\newcommand{\rem}[1]{}
\documentclass{amsart}
\usepackage{amsfonts,amssymb,amsmath,amsthm,mathrsfs}
\newtheorem{thrm}{Theorem}[section]
\newtheorem{remark}[thrm]{Remark}
\newtheorem{lem}[thrm]{Lemma}
\newtheorem{prop}[thrm]{Proposition}
\newtheorem{cor}[thrm]{Corollary}
\newtheorem{example}[thrm]{Example}
\theoremstyle{definition}
\newtheorem{definition}[thrm]{Definition}
\begin{document}  

\author[C.A.Mantica and L.G.Molinari]{Carlo Alberto Mantica and Luca Guido Molinari}
\address{C.~A.~Mantica and L.~G.~Molinari (corresponding author): 
Physics Department Aldo Pontremoli,
Universit\`a degli Studi di Milano and I.N.F.N. sezione di Milano,
Via Celoria 16, 20133 Milano, Italy.
ORCiD (L.G.Molinari): 0000-0002-5023-787X}
\email{carlo.mantica@mi.infn.it, luca.molinari@mi.infn.it}
\title[Doubly warped spacetimes]{Spherical doubly warped spacetimes\\
for radiating stars and cosmology}
\begin{abstract} 
Spherically symmetric spacetimes are ambient spaces for models of stellar collapse and
inhomogeneous cosmology. We obtain results for the Weyl tensor and the covariant form
of the Ricci tensor on general doubly warped (DW) spacetimes. In a spherically symmetric metric,
the Ricci and electric tensors become rank-2, built with a velocity vector field and its acceleration. 
Their structure dictates the general form of the energy-momentum tensor in the Einstein equations in DW spherical metrics. 
The anisotropic pressure and the heat current of an imperfect fluid descend from the gradient of the 
acceleration and the electric part of the Weyl tensor.\\
For radiating stellar collapse with heat flow, the junction conditions of the doubly warped metric with the Vaidya 
metric are reviewed, with the boundary condition for the radial pressure. 
The conditions for isotropy simply accomodate various models in the literature.\\
The anisotropy of the Ricci tensor in the special case of spherical GRW space-times (geodesic velocity), gives
Friedmann equations deviating from standard FRW cosmology by terms due to the electric tensor.\\
We introduce ``perfect 2-scalars" to discuss $f(R)$ gravity with anisotropic fluid source in a doubly warped spacetime, 
and show that the new geometric terms in the field equations
do not change the tensor structure of the fluid energy-momentum tensor.
\end{abstract}
\date{Nov. 2021, Jan. 2022}
\maketitle

\section{Introduction}
Spherical symmetry is a macroscopic property of many astrophysical objects and of the universe at large. 
For a static spherical mass, no matter what is the internal spherical distribution, the exterior is described by the Schwarzschild metric (1916), the unique solution of the Einstein equations that is asymptotically Minkowskian.
The Vaidya metric (1943) is the non static generalization that accounts for radiation emanating from a spherical mass that
changes in time \cite{Griffith}. 
In dimension $n\ge 4$ it is 
\begin{align}
 ds^2 =\left( 1-\frac{2m}{(n-3)r^{n-3}}\right)^{-1} \left [ - \frac{m_t^2}{m_r^2}dt^2+  dr^2 \right ] + r^2 d\Omega^2_{n-2}
 \label{Vaidya1}
 \end{align}
$m(t,r)$ is the total mass-radiation energy at time $t$ in a sphere of surface area $4\pi r^2$,
 $m_t$ and $m_r$ are partial derivatives \cite{Iyer89}\cite{Chatterjee90}. The interior metric solves the Einstein equations with a spherically symmetric source modelled
as an imperfect fluid. In proper coordinates, the metric is specified by three functions \cite{Tupper12}
$$ ds^2 = - B^2(t,r) dt^2 + C^2(t,r) dr^2 + D^2(t,r)d\Omega^2_{n-2} $$
Examples are the Lemaitre-Bondi-Tolman metric for the gravitational collapse of a sphere of
dust \cite{Bondi47}\cite{Gair01},  models of 
self-gravitating fluids \cite{Herrera97}\cite{Herrera04}, inhomogeneous cosmological models \cite{Bolejko}, neutron stars in extended gravity \cite{ACOO20}
\cite{ACOO21}\cite{Capozziello15}.
The functions satisfy conditions for continuity of first and second derivatives of the metric across
a timelike boundary manifold.

Because of the symmetry, the velocity field $u_j$ of the fluid is vorticity-free. 
It is also shear-free, i.e. $ \nabla_i u_j = \varphi (u_i u_j+g_{ij} ) - u_i \dot u_j $, 
if the ratio $C(t,r)/D(t,r)$ is independent of time \cite{Nairai68}\cite{Glass}. 
With this condition, the spacetimes are named by geometers (spherically symmetric) {\em doubly twisted}:
\begin{align}
 ds^2 = -b^2 (t,r) dt^2 + a^2(t,r)[ f_1(r)^2 dr^2 + f_2^2(r) d\Omega^2_{n-2} ]. \label{SDWmetric}
 \end{align}
The energy momentum tensor in the Einstein equations is often stated in the form
\begin{align}
T_{ij} = (\mu +p_\perp)u_i u_j +p_\perp
g_{ij} + u_iq_j + u_j q_i + (p_r-p_\perp) \chi_i \chi_j \label{EMT}
\end{align}
where $\chi_j $ is a unit vector parallel to $q_j$ and all functions depend on time $t$ and $r$.
The majority of studies are carried to the end in the simplified situation where the twist functions 
$b(t,r)$ and $a(t,r)$ factor. After a redefinition of time, the result is a (spherically symmetric) {\em doubly warped} metric:
\begin{align}
 ds^2 = -b^2 (r) dt^2 + a^2(t)[ f_1(r)^2 dr^2 + f_2^2(r) d\Omega^2_{n-2} ]. \label{sphere}
 \end{align}
They are still ambient spaces for models of stellar collapse and certain anisotropic cosmological models 
that will be considered here, and evolving wormholes \cite{Zangeneh14}\cite{Mehdizabeh21}.

 In this study we reverse the approach and first explore the geometry 
of such spacetimes and only then, turn to physics. 
The purpose is to clarify the context and the geometric nature of physics terms, in a covariant description. 

This is the structure ot the paper: in Section 2 we investigate general features of doubly twisted and doubly warped (DW) spacetimes, with statements
about the Weyl tensor, the Ricci tensor, and the space submanifold.\\
In Section 3 we specialize to spherically symmetric DW spacetimes. 
We show that the unit radial vector is torse-forming, and obtain the covariant expressions of the Ricci tensor 
and of the electric component of the Weyl tensor, with a useful identity relating the Ricci tensor of the space submanifold to the electric tensor. \\ 
In Section 4 the structure of the Ricci tensor, here written with functions $R_i(t,r)$,
$$ R_{ij} = R_1 u_i u_j + R_2 g_{ij} + R_3 ( u_i \dot u_j + u_j \dot u_i ) + 
 R_4 \left (\frac{\dot u_i \dot u_j}{\dot u^k\dot u_k}- \frac{h_{ij}}{n-1} \right ) $$
entrains a similar one for the energy-momentum tensor $T_{ij}$. 
The pressure anisotropy is encoded in $R_4$ that results from the electric tensor and the gradient of the acceleration $\dot u_j$. 
The boundary conditions connecting the general spherical DW metric to the Vaidya metric are reviewed
along with the prescriptions by Santos \cite{Santos85}.
The parameters of the inner metric determine the mass function $m(t,r)$ and a differential equation 
for the inner scale function $a(t)$. The latter has the physical meaning that, at the boundary, the radial pressure equals the 
modulus of the heat current. \\
In Section 5 we study isotropic DW spherical solutions. The isotropy
condition $R_4=0$ determines equations for the fields $b$, $f_1$ and $f_2$ in \eqref{SDWmetric}, that provide a frame for models of stellar collapse studied by various authors.\\
In Section 6 we consider the spherical spacetimes \eqref{SDWmetric} with $b(r)=1$ (equivalent to $\dot u_i=0$). For such Generalized Robertson-Walker (GRW) metric
we obtain the analogue of the Friedmann equations, where pressure is anisotropic because of the electric tensor. \\
In Section 7 we introduce perfect 2-scalars, in connection with the presence of two distinguished vectors, $u_i$ and $\dot u_i$ that build the
relevant tensors. They extend the fruitful concept of perfect 1-scalars introduced by us in connection with a rank-1 Ricci tensor (quasi-Einstein or perfect fluid) in RW spacetimes, with a single distinguished vector \cite{Capozziello2022}.
We prove that, in $f(R)$ gravity with anisotropic fluid source, the geometric corrections have the same tensor form of 
the fluid source.\\
Sections 8 and 9 contain conclusions and appendixes.

We employ Latin letters for spacetime indices and Greek letters for space ones. A dot is the directional derivative along 
$u_i$: $\dot X=u^k\nabla_k X$; for $b\neq 1$ it does not coincide with the time derivative in the comoving frame. 
To facilitate the reading, we postponed to appendixes some long or marginal proofs.

\section{Doubly twisted and doubly warped spacetimes}
Doubly twisted spacetimes are a large class of Lorentzian manifolds introduced by Kentaro Yano in 1940 \cite{Yano24}. He showed that the metric structure\footnote{Some authors exchange the letters $a^2$ and $b^2$. Here $a^2$ is minding the scale function of RW metric, a limit case in this class.} 
\begin{align}
 ds^2 = -b^2(t,{\bf x}) dt^2 +a^2(t,{\bf x})  g^\star_{\mu\nu}({\bf x}) dx^\mu dx^\nu \label{1.0}
 \end{align}
is necessary and sufficient for the spatial foliation to be totally umbilical. 
Hereafter, the Riemannian spatial submanifold ($M^\star, g^\star$) has dimension $n-1$. \\
In physics the best known examples are the Stephani universes \cite{Stephani}\cite{Krasinski}. They are conformally flat solutions of the Einstein equations with perfect fluid source ($n=4$),
$$ds^2 = - b^2(t, {\bf x})dt^2 + R^2(t) (dx^2 + dy^2 + dz^2)/ V^2(t,{\bf x}) $$
with $V = 1 +\frac{1}{4}\| {\bf x} - {\bf x}_0(t)\|^2$, $b^2 = F(t) [\dot V /V - \dot R /R ]$. 
Another example is the conformally flat solution for a perfect fluid with heat flux by Banerjee et al. \cite{Banerjee89}:
$$ds^2 = - V^2 (t,{\bf x}) dt^2 + (dx^2 +dy^2 +dz^2)/U^2(t,{\bf x})$$
with $ UV = A(t) \| {\bf x} \|^2+ {\bf A}(t)\cdot {\bf x} + A_0(t)$, $U = B(t)\|{\bf x}\|^2+ (t) \cdot {\bf x} + B_0(t)$, where the functions of time are arbitrary.

As shown in \cite{Mantica21}, doubly twisted spacetimes may be covariantly characterised by the existence of a {\em time-like doubly torqued} vector field: 
$\tau_i\tau^i<0$ and 
 \begin{align}
\nabla_j \tau_k = \kappa g_{jk} +\alpha_j \tau_k + \tau_j \beta_k \label{2.1}
\end{align}
with vector fields $ \alpha_j \tau^j=0$ and $\beta_j \tau^j=0$.
%The vector $\tau_i$ is hypersurface orthogonal, i.e. there are scalar functions
%$\sigma $ and $F$ such that $\tau_j (x) =\sigma (x) \partial_j F(x)$. 
%Then:$ \partial_i \tau_j = \frac{\partial_i \tau}{\tau} \tau_j + \tau \partial_i\partial_j F $.
Depending on $\alpha_i $, $\beta_i$ and $\kappa $, a classification of doubly twisted spacetimes results
(Table 1 in \cite{Mantica21b}) that includes twisted, doubly warped (DW), generalized Robertson-Walker (GRW) and
static spacetimes.\\
In the {\em comoving frame}, defined by space components $\tau_\mu =0$, 
the spacelike vectors $\alpha_i $ and $\beta_i $ are in simple relation 
to the functions $a$ and $b$ in the metric \eqref{1.0} \cite{Mantica21}:
 \begin{align}
 \begin{array}{l} \alpha_0=0\\ 
 \alpha_\mu = \partial_\mu \log a \end{array} \qquad 
 \begin{array}{l} \beta_0=0\\
 \beta_\mu = -\partial_\mu \log b \end{array} \qquad 
 \kappa =\frac{\partial_t a}{b}  \label{2.3}
 \end{align}
% A rescaled vector $\tau'_j = \lambda \tau_j$ satisfies 
%$\nabla_j \tau'_k = (\lambda\kappa) g_{jk} +(\alpha_j + \partial_j\log %\lambda )\tau'_k + \tau'_j \beta_k $. 
%It is still doubly torqued if $\tau^j\partial_j \lambda =0$, i.e. $\partial_t \lambda =0$
%in the coordinates \eqref{1.0}. Then, by eq.\eqref{2.3}, it is $a'({\bf x},t)= \lambda ({\bf x}) a({\bf x},t) $.  

An equivalent description 
is in terms of the time-like unit vector field \cite{Ferrando}
$$u_j = \frac{\tau_j}{\sqrt{-\tau^2}} , \quad  u^j u_j =-1. $$
%Though a rescaling of $\tau_i$ with $\lambda =1/\sqrt{-\tau^2}$, the vector is not doubly torqued because $\tau^i \partial_i \lambda $ is non-zero. % (it is $\tau^i\nabla_i\tau^2 =2\kappa \tau^2$).
Eq.\eqref{2.1} becomes $\nabla_i u_j =\varphi ( g_{ij}+u_iu_j) + u_i\beta_j $, with $\varphi =\kappa/\sqrt{-\tau^2}$. The contraction with $u^i$ gives $\beta_j = -u^k\nabla_k u_j=-\dot u_j$.  
Therefore, the second characterization of doubly twisted spacetimes is:
\begin{align}
 \nabla_i u_j = \varphi ( g_{ij}+u_iu_j)- u_i\dot u_j .
\end{align}
The following result introduces doubly warped spacetimes:

\begin{prop}\label{closedalpha}
The vector $\alpha_i $ in eq.\eqref{2.1} is closed if and only if $\nabla_i\varphi = -u_i \dot\varphi - \varphi \dot u_i $.
\begin{proof}
The gradient of $\varphi =\kappa/\sqrt{-\tau^2}$ and the identity $\nabla_i \tau^2 =2\kappa \tau_i + 2\alpha_i \tau^2$ 
give 
\begin{align}
 \frac{\nabla_i \varphi }{\varphi} - \varphi u_i = \frac{\nabla_i  \kappa }{\kappa } -\alpha_i  \label{PHIKAPPA}
 \end{align}
Another gradient and antisymmetrization in the indices give
$$\nabla_i\alpha_j -\nabla_j\alpha_i = -\varphi (u_i\dot u_j -u_j\dot u_i)+u_j \nabla_i\varphi -u_i\nabla_j\varphi   $$
If $\alpha_i$ is closed, contraction with $u^i$ gives the property. The opposite is also true.
\end{proof}
\end{prop}
\rem{RRRRRRRRR
\begin{remark}\label{CorClosed}
If $\alpha_i$ is closed ($\alpha_i=\partial_i \alpha $) then  $\varphi u_i$ (=$\kappa\tau_i/\tau^2$) is closed and hypersurface orthogonal: 
$\varphi u_i = \partial_i F$, and $F =\log(\varphi \kappa) + \alpha $.\\
A consequence of Prop.2.2 in ref.\cite{Mantica21} is:
if $\alpha_i$ is closed and $\nabla_i\kappa =0$ then $\alpha_i +\beta_i=0$.
\end{remark}
RRRRRRR}

\subsection{Doubly warped (DW) space-times} 
A doubly warped space-time is a special case of the metric \eqref{1.0} where  $b^2$ is independent of time and $a^2$ is independent of position:
\begin{align}
 ds^2 = -b^2({\bf x}) dt^2 +a^2(t)  g^\star_{\mu\nu}({\bf x}) dx^\mu dx^\nu \label{1.1}
 \end{align}
We present three equivalent covariant characterisations. Two of them specialize the previous ones. We shall exploit the one based on a time-like unit vector.
\subsection*{$\bullet $ Doubly torqued vector} (Mantica, Molinari \cite{Mantica21}\cite{Mantica21b})
{\em A Lorentzian manifold is doubly-warped if and only if there is a time-like doubly torqued vector field \eqref{2.1} 
with $\alpha_j $ and $\beta_j $ closed.}

With the metric \eqref{1.1} the relations \eqref{2.3} show that $\alpha_i=0$, $\beta_0=0$ and $\beta_\mu =-\partial_\mu \log b({\bf x})$. Then $\beta_i$ is closed. One can relax the condition $\alpha_i=0$ to $\alpha_i$ closed by noting that
a rescaled vector $\tau'=\lambda \tau$ 
%satisfies $\nabla_j \tau'_k = (\lambda\kappa) g_{jk} +(\partial_j\log \lambda )\tau'_k + \tau'_j (-\partial_k \log b)$. It 
is still doubly torqued if $\tau^i\partial_i\lambda =0$ i.e. $\lambda $ does not depend on time in the frame \eqref{1.1}. 
The metric changes in the factor $a'(t,{\bf x}) = a(t)\lambda ({\bf x})$, which amounts to a conformal change of the inner metric $g^\star ({\bf x})$ of the submanifold. 

\subsection*{$\bullet $ Conformal Killing vector} (Ramos et al. \cite{Ramos}, Th.1): 
{\em A space-time is doubly-warped if and only if there exists a non-null, nowhere zero and hypersurface 
orthogonal vector field $\zeta_i$ that is conformal Killing, $\pounds_\zeta g_{ij} = 2\kappa g_{ij}$, 
with $\nabla_j \kappa = \sigma \zeta_j$ for some nonzero scalar function $\sigma $}. The equivalence is proven in the next
Prop.~\ref{PKILL}.

\subsection*{$\bullet $ Timelike unit vector}
{\em A Lorentzian manifold is doubly-warped if and only if there is a time-like unit vector field such that:}
\begin{align}
&\nabla_i u_j = \varphi (u_i u_j + g_{ij}) - u_i \dot u_j  \label{U1}   \\
&\nabla_j \varphi = -u_j \dot \varphi  - \varphi \dot u_j \label{U2}\\
&\nabla_i \dot u_j = \nabla_j \dot u_i \label{U3}
\end{align}
Everything follows from Prop.~\ref{closedalpha} and the identification $\beta_i =-\dot u_i$, with $\beta_i$ closed.\\
- If $\varphi =0$ then $\nabla_i u_j =  - u_i \dot u_j$ and $\nabla_i \dot u_j = \nabla_j \dot u_i$: 
this is the definition of static space-time (\cite{Stephani}, p.283).\\
- If $\dot u_j=0$ then $\nabla_i u_j = \varphi (u_i u_j + g_{ij}) $ with $\nabla_j \varphi = -u_j \dot \varphi $: this is a generalized Robertson-Walker spacetime 
\cite{Mantica17surv}. Equivalently: $b=1$ in the metric \eqref{1.1}.

\begin{lem} In a DW spacetime, $\varphi u_i $ is closed and orthogonal to hypersurfaces of constant time.
\begin{proof}
$\nabla_i(\varphi u_j) = \varphi [\varphi (u_iu_j+g_{ij}) -u_i\dot u_j]
+u_j (-u_i \dot\varphi - \dot u_i\varphi) = (\varphi^2-\dot\varphi)u_iu_j+\varphi^2 g_{ij} -\varphi (u_i\dot u_j + u_j\dot u_i)$.
Then $\nabla_i(\varphi u_j)=\nabla_j(\varphi u_i)$ and $\varphi u_j =\partial_j F$ (a gradient). In the comoving frame: 
$u_0=-b$, $u_\mu =0$  and (see Appendix \ref{DWcoeffs})
\begin{align}
\varphi = \frac{\nabla_j u^j}{n-1} = \frac{1}{n-1} \Gamma_{\mu 0}^\mu u^0 =  \frac{a_t}{ab}
\end{align}
Then $F=-\log  a(t)$ and $u_i$ is orthogonal to hypersurfaces of constant time.
\end{proof}
\end{lem}

\begin{prop}\label{PKILL} In a DW spacetime there exists a scalar function $\theta $ such that 
$$ \nabla_i (\theta u_j) = (\theta\varphi ) g_{ij} + \dot u_i (\theta u_j) - (\theta u_i)\dot u_j $$
Then $\theta u_i$ is a conformal Killing vector (and a torse-forming vector).
\begin{proof}  
The conformal Killing condition $\pounds_{\theta u}g_{ij}= \nabla_i(\theta u_j)+\nabla_j(\theta u_i) =2 \kappa g_{ij}$ is satisfied if and only if $\kappa =\theta\varphi $ and $u_i(\nabla_j\theta +\varphi\theta u_j -\theta \dot u_j) +u_j(\nabla_i\theta +\varphi\theta u_i -\theta \dot u_i) =0$
i.e. $\nabla_i \log\theta = -(\varphi u_i) +\dot u_i$. The function $\theta $ exists because both vectors $\varphi u_i$ and $\dot u_i$ are gradients.  It is also $ \nabla_i (\theta\varphi) = - (\varphi^2 +\dot\varphi) (\theta u_i) $ (the homothetic case $\varphi^2+\dot\varphi =0$ is not considered here).
\end{proof}
\end{prop}

\subsection{Special space metrics $g^\star $ in DW spacetimes} 
The presence of the distinguished spacelike vector $\dot u_i$ allows for a classification of the metrics $g^\star $
of the space sub-manifold. The gradient of the vector is decomposed in its components along $u_i$, $\dot u_i$,
and orthogonal ones, through the projector
\begin{align}
N_{jk} = g_{jk}+u_ju_k -\frac{1}{\eta}\dot u_j \dot u_k, \qquad \eta \equiv \dot u_j \dot u^j \label{projector}
\end{align}
\begin{prop} \label{PROP_nabladotu}
On a doubly warped spacetime: 
\begin{align}
\nabla_j \dot u_k =& -\eta u_j u_k +\varphi (u_j \dot u_k + \dot u_j u_k) + \dot u_j \dot u_k \frac{\dot u^i\nabla_i\eta}{2\eta^2}
+ N_{jk} \frac{N_{rs} \nabla^r \dot u^s}{n-2} \label{nabladot}\\
&+\dot u_j w_k + w_j \dot u_k + \Pi_{jk}\nonumber\\
w_k=&\frac{1}{2\eta}N_{km}\nabla^m\eta \\
 \Pi_{jk}=& \left [N_{jr}N_{ks}-  \frac{N_{jk} N_{rs}}{n-2} \right]\nabla^r\dot u^s \\
\nabla_j \eta =& 2\varphi \eta u_j +  \dot u_j  \frac{\dot u^i\nabla_i\eta}{\eta} + 2\eta w_j \label{nablaeta}
\end{align}
where  $w_k$ is orthogonal to $u_j$ and $\dot u_j$, 
$\Pi_{ij}$ is traceless symmetric and annihilates 
$u_i$ and $\dot u_i$. Both $w_k$ and $\Pi_{ij}$ are spacelike tensors. 
\begin{proof}
The procedure is laborious but standard \cite{Clarkson}. Write $\nabla_j \dot u_k = g_{jr}g_{ks}\nabla^k \dot u^s$ and replace the metric tensor $g_{jr}$ with $N_{jr} - u_j u_r + \frac{1}{\eta}\dot u_j \dot u_r$ and similarly for $g_{ks}$. 
Use $u^i\dot u_i=0$ and the symmetry of $\nabla_i\dot u_j$ to simplify terms: 
$u_r \nabla^r \dot u^s = u_r \nabla^s\dot u^r = - \dot u_r \nabla^s u^r = -\varphi\dot u^s + u^s\eta $. 
Then: $u^r\nabla_r \eta = 2u^r\dot u^s\nabla_r \dot u_s= -2\varphi \eta$.
\end{proof}
\end{prop}
%
%\begin{cor}
%In a doubly warped spacetime:
%\begin{align} 
%&u^j\nabla_j \dot u_i =  -\varphi \dot u_i+\eta u_i \\
%&u^j\nabla_j \eta = -2\varphi \eta
%\end{align}
%\end{cor}
%
Now consider the decomposition \eqref{nabladot} in the comoving frame, where $u_\mu=0$. There it is $\dot u_0=0$, and $\dot u_\mu$ are the space components. Since $\eta =  g^{\star\mu\nu} ({\bf x}) \dot u_\mu \dot u_\nu /a^2(t) $, the vector field
\begin{align}
 {\hat n}_\mu = \frac{\dot u_\mu}{a(t)\sqrt \eta} \label{enne}
 \end{align}
 is a unit vector field on $M^\star$: $1=g^{\star\mu\nu} {\hat n}_\mu {\hat n}_\nu $. The space components of the decomposition \eqref{nabladot} give:
\begin{prop}\label{nablaenne}
\begin{gather}
\nabla^\star_\mu {\hat n}_\nu = \frac{\Theta}{n-2}(g^\star_{\mu\nu} - {\hat n}_\mu {\hat n}_\nu) + {\hat n}_\mu w_\nu + \Pi^\star_{\mu\nu} \\
\Theta = \frac{a}{\sqrt\eta} N_{rs}\nabla^r \dot u^s=\nabla^\star_\mu {\hat n}^\mu , \qquad \Pi^\star_{\mu\nu} = \frac{\Pi_{\mu\nu}}{a\sqrt \eta}\nonumber
\end{gather}
%Some special cases are now given.  Let ${\bf x}=(x,{\bf y})$ be the space coordinates adapted to the direction 
%$n_j$.
$\bullet$ If $\Pi^\star_{\mu\nu}=0$ the space submanifold is doubly twisted 
(\cite{Ferrando} eq.4, \cite{Borowiec} table 1, \cite{Eisenhart} eq. 53.14). There are coordinates ${\bf x}=(x,{\bf y})$ and functions 
$f_1$ and $f_2$ such that
$$g^\star_{\mu\nu} dx^\mu dx^\nu = f_1^2(x,{\bf y})dx^2 + f_2^2(x,{\bf y}) g^{\star\star}({\bf y})_{qp} dy^q dy^p$$
$\bullet$ If $\Pi^\star_{\mu\nu}=0$ and $w_\mu =0$ the space submanifold is twisted \cite{Mantica17}\cite{Tacchini}. 
There are coordinates and a function $f$ such that
$$g^\star_{\mu\nu} dx^\mu dx^\nu = dx^2 + f^2(x,{\bf y}) g^{\star\star}({\bf y})_{qp} dy^qdy^p$$
$\bullet$ If $\Pi^\star_{\mu\nu}=0$, $w_\mu =0$ and $\partial_\mu \Theta = {\hat n}_\mu {\hat n}^\nu\partial_\nu\Theta$ the space submanifold is warped \cite{ManticaJMP}. There are coordinates and a function 
$f$  such that
$$g^\star_{\mu\nu} dx^\mu dx^\nu = dx^2 + f^2(x) g^{\star\star}({\bf y})_{qp} dy^qdy^p$$
The vector ${\hat n}_\mu$ is an eigenvector of the Ricci tensor with eigenvalue $\xi^\star $. The integrability 
condition and the Weyl tensor of the space submanifold give the Ricci tensor\footnote{the proof is analogous to that
of \eqref{RicciRicci}, with the difference that the metric $g^\star$ is Riemannian.}.
\begin{align}
R^\star_{\mu\nu}= \xi^\star {\hat n}_\mu {\hat n}_\nu  + \frac{R^\star- \xi^\star}{n-2}(g^\star_{\mu\nu}- {\hat n}_\mu {\hat n}_\nu )+
(n-3) {\hat n}^\tau {\hat n}^\sigma  C^\star_{\tau\mu\nu\sigma}   \label{Riccistar}
\end{align}
where $R^\star = g^{\star \mu\nu}R^\star_{\mu\nu}$.
\end{prop}

\subsection{The Weyl tensor on doubly warped spacetimes} 
Kentaro Yano (1957, \cite{Yano57} eq.~3.8 page 161) proved that if  $v$ is a proper conformal vector field, then:\\
1) the Lie derivative along $v$ of the Weyl tensor is zero,
\begin{align*}
\pounds_v C_{jkl}{}^m =0, 
\end{align*}
2) the Lie derivative of the divergence of the Weyl tensor is (see \cite{Sharma88}):
\begin{align}
\pounds_v (\nabla_m C_{jkl}{}^m ) =(n-3)( \nabla_m \kappa) C_{jkl}{}^m. \label{Lienabla}
\end{align}
In Prop.~\ref{PKILL} we showed that a DW space-time is always endowed with a proper conformal Killing vector $\theta u_i$. Therefore, the above two 
properties hold true in DW spacetimes. The second one gives an interesting result.\\ 
First note that being $u_i$ proportional to a doubly torqued vector, it is Weyl compatible
(the proof for doubly torqued vectors is given in  \cite{Mantica21b}):
\begin{align}
u_i C_{jklm}u^m +u_j C_{kilm}u^m +u_k C_{ijlm}u^m =0.  \label{WeylCu}
\end{align}
%
%\begin{lem} If $\tau $ is a proper conformal vector field, then the Lie derivative of the divergence of the Weyl tensor is:
%\begin{align}
%\pounds_\tau (\nabla_m C_{jkl}{}^m ) =(n-3)( \nabla_m \kappa) C_{jkl}{}^m
%\end{align}
%\begin{proof}
%Consider the commutator:
%\begin{align*}
%\pounds_\tau (\nabla_iC_{jkl}{}^m) - &\nabla_i \pounds_\tau C_{jkl}{}^m = (\pounds_\tau \Gamma_{ip}^m)C_{jkl}{}^p 
%\label{pound1}\\
%&-(\pounds_\tau \Gamma_{ij}^p) C_{pkl}{}^m 
%-(\pounds_\tau \Gamma_{ik}^p) C_{jpl}{}^m -(\pounds_\tau \Gamma_{il}^p) C_{jkp}{}^m \nonumber
%\end{align*}
%With the Lie derivative of the Christoffel symbols $\pounds_\tau \Gamma_{jk}^i =\delta_k^i \nabla_j \kappa  +\delta_j^i %\nabla_k\kappa - g_{jk} \nabla^i\kappa $ (see \cite{Yano57} page 161) and \eqref{PO2}, the equation  becomes eq.3.13 in \cite{Yano57} page 162: 
%\begin{align}
%\pounds_\tau (\nabla_iC_{jkl}{}^m) = -2\kappa_i C_{jkl}{}^m +\delta_i^m \kappa_p C_{jkl}{}^p -\kappa^m C_{jkli} -
%\kappa_j C_{ikl}{}^m - \kappa_k C_{jil}{}^m -\kappa_l C_{jki}{}^m \nonumber\\
%+g_{ij}\kappa^p C_{pkl}{}^m + g_{ik} \kappa^p C_{jpl}{}^m + g_{il} \kappa^p C_{jkp}{}^m 
%\end{align}
%where $\kappa_j=\nabla_j \kappa $. The contraction with $\delta^i_m$ gives the result.
%\end{proof}
%\end{lem}.
%
\begin{prop}\label{Cu} On a doubly warped spacetime of dimension $n$:
\begin{gather}
\nabla_m C_{jkl}{}^m=0 \quad \Longrightarrow \quad C_{jklm} u^m =0\quad \text{and}\quad C_{jklm} \dot u^m =0, \quad (n\ge 4)\\
\nabla_m C_{jkl}{}^m =0 \quad \Longrightarrow \quad C_{jklm}=0 \qquad (n=4)
\end{gather}
\begin{proof}
If $\nabla_m C_{jkl}{}^m=0$, then $\pounds_{\theta u}  \nabla_m C_{jkl}{}^m=0$. According to eq.\eqref{Lienabla} it is
$$0= \nabla_m (\theta\varphi)C_{jkl}{}^m =-(\varphi^2+\dot\varphi) C_{jkl}{}^m  (\theta u_m)$$ 
Then, if $\varphi^2+\dot\varphi \neq 0$: $C_{jkl}{}^m  u_m=0$. \\
%If $\varphi^2+\dot\varphi = 0$, as shown in Remark \ref{HOMO}, the spacetime is GRW. Then it is known that
%$\nabla_m C_{jkl}{}^m=0$ implies $C_{jklm}u^m=0$.
With $\nabla_m C_{jkl}{}^m=0$ the second Bianchi identity for the Weyl tensor \cite{Adati} has the simpler form 
$$\nabla_i C_{jklm}+\nabla_j C_{kilm} + \nabla_k C_{ijlm}=0.$$
The contraction with $u^m$ and the relation $(\nabla_i C_{jklm})u^m = 
C_{jklm}\nabla_i u^m$  give:
\begin{align*}
 0 =&-C_{jklm}  \nabla_i u^m - C_{kilm}\nabla_j u^m - C_{ijlm}\nabla_k u^m \\
 =& u_i C_{jklm} \dot u^m + u_jC_{kilm}\dot u^m + u_k C_{ijlm}\dot u^m  -\varphi (C_{jkli}+ C_{kilj}+C_{ijlk})
 \end{align*}
The contraction with $u^i$ and the first Bianchi identity conclude the first proof.\\
The spacetime dimension $n=4$ is special: Lovelock proved that 
$C_{jklm}u^m=0$ for a non-null vector implies 
$C_{jklm}=0$ (see \cite{LovelockRund}). 
\end{proof}
\end{prop} 
The electric tensor $E_{kl} =u^ju^m C_{jklm}$ is symmetric, traceless and $E_{ij}u^j=0$. We'll show that it appears in the expression of the Ricci tensor. The property 
\eqref{WeylCu} implies $C_{jklm}u^m = u_k E_{jl}- u_j E_{kl}$. Then $C_{jklm}u^m =0  \Longleftrightarrow E_{jk}=0$, and

\begin{cor} $\nabla_m C_{jkl}{}^m =0 \quad \Longrightarrow\quad  E_{ij}=0. $
 \end{cor}

We conclude the section with quotations:\\
$\bullet$ In a DW spacetime if $\nabla_i C_{jklm}=0$ then $C_{jklm}=0$ (Hotlo\'s \cite{Hotlos}, Th.6).\\ 
%Every conformally symmetric $(\nabla_i C_{jklm}=0)$ DW manifold is conformally flat $(C_{jklm}=0)$ (Hotlo\'s \cite{Hotlos}, Th.6).\\ 
$\bullet$ In a DW spacetime if $C_{jklm}=0$ then $C^\star_{jklm}=0$ (G\c{e}barowski \cite{Gebarowski95}).\\
$\bullet$ In a GRW spacetime ($b=1$ i.e. $\dot u_i=0$): $\nabla_m C_{jkl}{}^m=0 \Longleftrightarrow C_{jklm}u^m=0$
\cite{ManticaJMP}.\\
$\bullet$ Banerjee \cite{Banerjee81} proved that 
$ds^2=-b^2({\bf x},t) dt^2 + a^2({\bf x},t)\delta_{\mu\nu} dx^\mu dx^\nu $ 
is conformally flat ($C_{jkl}{}^m=0$) if and only if
$b({\bf x},t) = a({\bf x},t) [A(t) {\bf x\cdot  x} + (t){\bf \cdot x}   + C(t) ]$
for any $A(t)$, $ (t)$ and $C(t)$. 
As a special case we have:
\begin{prop}
A doubly warped spacetime  $ds^2=-b^2({\bf x}) dt^2 + a^2(t)\delta_{\mu\nu} dx^\mu dx^\nu $
is conformally flat  if and only if $b(r) = K r^2  + C$, with constants $K$,  $C$.
\end{prop}

\subsection{The Ricci tensor on doubly warped spacetimes}
We obtain the general covariant expression of the Ricci tensor, and a useful identity among the Ricci tensor $R^\star_{\mu\nu}$ of the
space submanifold and the electric tensor.
\begin{prop} The Ricci tensor on a doubly warped spacetime is
\begin{align}
R_{jk} =& u_j u_k \left[ \frac{R-n\xi}{n-1} + \nabla^p\dot u_p +\eta + \frac{\dot u^s\nabla_s \eta}{2\eta} \right ]
+ g_{jk} \left[ \frac{R-\xi}{n-1} + \eta + \frac{\dot u^s\nabla_s \eta}{2\eta} \right ] \nonumber\\ 
& - (n-2)\varphi (u_j \dot u_k + \dot u_j u_k) - \frac{\dot u_j \dot u_k}{\eta} 
\left [(n-1) \left ( \eta + \frac{\dot u^i\nabla_i\eta}{2\eta} \right ) -\nabla^p\dot u_p \right ] \label{RicciRicci}\\
&-(n-2)[\dot u_j w_k + w_j \dot u_k + \Pi_{jk} + E_{jk}]\nonumber
\end{align} 
where $\xi = (n-1)(\varphi^2+\dot\varphi) $, $\eta = \dot u_p \dot u^p$ and $E_{kl}=u^j u^m C_{jklm}$ (electric tensor). 
The vector $w_j$ and the tensor $\Pi_{ij}$ were given in Prop.\ref{PROP_nabladotu}.
\begin{proof}
The integrability condition $R_{jkl}{}^m u_m = [\nabla_j,\nabla_k]u_l$ with \eqref{U1} becomes:
\begin{align*}
R_{jklm} u^m = (\varphi^2+\dot\varphi)(u_k g_{jl}-u_j g_{kl})-\varphi (\dot u_j g_{kl}-\dot u_k g_{jl})\\ +
u_j (\dot u_k\dot u_l + \nabla_k \dot u_l) -u_k (\dot u_j\dot u_l + \nabla_j \dot u_l)  \nonumber
\end{align*}
The contraction with $g^{jl}$ is
$R_{km} u^m = (\xi-\eta - \nabla_p \dot u^p ) u_k +(n-1)\varphi \dot u_k  +
u^j  \nabla_k \dot u_j  $. 
With the simplification $u^j  \nabla_k \dot u_j =-\dot u^j \nabla_k u_j = -\varphi \dot u_k + \eta u_k $, 
the result is:
\begin{align}
R_{km} u^m = (\xi - \nabla^p\dot u_p ) u_k +(n-2)\varphi \dot u_k \label{Ricciu}
\end{align}
An expression for the Ricci tensor is now obtained with the Weyl tensor
\begin{align*}
C_{jklm} = R_{jklm}+\frac{g_{jm} R_{kl}-g_{km}R_{jl} +R_{jm}g_{kl}-R_{km} g_{jl}}{n-2} - R\frac{g_{jm}g_{kl}-g_{km}g_{jl}}{(n-1)(n-2)}
\end{align*}
The contraction with $u^ju^m$ and the previous relations give:
\begin{align*}
E_{kl}=& (\varphi^2+\dot\varphi)(u_k u_l + g_{kl}) - \dot u_k\dot u_l - \nabla_k \dot u_l -u_k (u^j \nabla_j \dot u_l)
-\varphi u_k \dot u_l \\
&-\frac{R_{kl} +(\xi-\nabla^p\dot u_p)(g_{kl}+2u_k u_l)}{n-2}+ R\frac{g_{kl}+u_k u_l}{(n-1)(n-2)}. 
\end{align*}
The Ricci tensor is obtained:
\begin{align*}
R_{kl} =& u_k u_l \left[ \frac{R-n\xi}{n-1} + 2\nabla_p\dot u^p  \right ] + g_{kl} \left[ \frac{R-\xi}{n-1} + \nabla_p\dot u^p  \right ] \\
& - (n-2)(\nabla_k \dot u_l +\dot u_k \dot u_l + u_k \ddot u_l +\varphi u_k \dot u_l +E_{kl}) 
\end{align*}
Note that $\ddot u_l = u^j\nabla_j \dot u_l =  u^j\nabla_l \dot u_j = -\dot u^j \nabla_l u_j = -\varphi \dot u_l +u_l\eta $. This simplifies the last line.
Next insert the decomposition \eqref{nabladot} for $\nabla_k \dot u_l$:
\begin{align*}
R_{kl} =& u_k u_l \left[ \frac{R-n\xi}{n-1} + 2\nabla_p\dot u^p  -N_{rs} \nabla^r\dot u^s \right ] 
+ g_{kl} \left[ \frac{R-\xi}{n-1} + \nabla_p\dot u^p  -N_{rs} \nabla^r\dot u^s  \right ] \\
& -(n-2)\varphi (u_k \dot u_l +\dot u_k u_l) 
 - (n-2)\frac{\dot u_k \dot u_l }{\eta}\left [ \eta + \frac{u^p\nabla_p \eta}{2\eta} - \frac{N_{rs}\nabla^r \dot u^s}{n-2}\right]\\
 & - (n-2)(\dot u_k w_l + \dot u_l w_k + \Pi_{kl} +E_{kl})
\end{align*}
Finally, specify the term
\begin{align} 
N_{rs}\nabla^r \dot u^s=\nabla_r \dot u^r + u^r u^s \nabla_r \dot u_s - \frac{\dot u_r \dot u_s}{\eta} \nabla^r \dot u^s
=\nabla_p\dot u^p -\eta -\frac{\dot u^s\nabla_s\eta}{2\eta}  \label{Nnabladot}
\end{align}
The expression \eqref{RicciRicci}  is now obtained.
\end{proof}
\end{prop}
The electric tensor is spacelike. In the comoving frame  \eqref{1.1} where $u^\mu =0$, it is $E_{00}=E_{0\mu}=0$. The evaluation of the space components $E_{\mu\nu}$ (in Appendix \ref{EVAL_EMUNU}) provides the following relation with $R^\star_{\mu\nu}$, the Ricci tensor on ($M^\star, g^\star $):
\begin{prop}\label{3.12} On a doubly warped spacetime:
\begin{align}
R^\star_{\mu\nu} - \frac{R^\star}{n-1}g^\star_{\mu\nu} =  -(n-3) \frac{1}{b} \left[ \nabla^\star _\mu b_\nu - \frac{\nabla^\star_\rho b^\rho }{n-1} g^\star_{\mu\nu} \right ]-(n-2) E_{\mu\nu}. \label{GebaExt}
\end{align}
where $b_\nu  =\partial_\nu b$ and $b^\mu = g^{\star \mu\nu}\partial_\nu b$, $R^\star = g^{\star \mu\nu}R^\star_{\mu\nu}$.
\end{prop}
\noindent
The identity extends a result obtained by G\c{e}barowski  
(\cite{Gebarowski84}, eq.23) under the restriction of harmonic Weyl tensor ($\nabla_m C_{jkl}{}^m=0$ i.e. $E_{\mu\nu}=0$). It will be very
useful in conjunction with spherical symmetry.
\begin{prop}\label{DOTRetc}
\begin{gather}
\nabla_k \xi= -u_k \dot \xi -2\dot u_k \xi \label{nablaxi} \\
u^k\nabla_k(\nabla_p\dot u^p) =-2\varphi (\nabla^p\dot u_p) \label{dotnabladotu}\\
\dot R  -2\dot \xi =-2\varphi (R -n\xi) \label{dotR}
\end{gather}
\end{prop}
\noindent
The proofs are in Appendix \ref{PROOFDOTR}. 
Remarkably, eq.\eqref{dotR} has the same simple form as in GRW and RW spacetimes \cite{Capozziello2022}. 
Its solution eq.\eqref{RRstar} can be written as
\begin{align}
R=\frac{R^\star}{a^2}  - 2\nabla_j \dot u^j + 2\xi + (n-1)(n-2)\varphi^2.  \label{Rcov}
\end{align}

\subsection{The Ricci tensor on GRW spacetimes}
On GRW spacetimes $(\dot u_i=0)$ the Ricci tensor is \cite{Mantica17surv}:
\begin{align}
R_{jk} = u_j u_k \frac{R-n\xi}{n-1}+ g_{jk}  \frac{R-\xi}{n-1} - (n-2) E_{jk} \label{RicciGRW}
\end{align}
Eq.\eqref{GebaExt} shows that the Ricci tensor is perfect fluid ($E_{jk}=0$) if and only if $R^\star_{\mu\nu}$ is Einstein.
The Bianchi identity, eq.\eqref{nablaxi} (with $\dot u_i=0$) and  \eqref{dotR} give  
\begin{align}
\nabla_k R =- u_k\dot R   -\frac{2(n-1)(n-2)}{n-3}\nabla^j E_{jk} \label{divRicciGRW}
\end{align}
By eq.\eqref{GebaExt} the spacelike vector $\nabla^j E_{jk}$ has space components proportional to $\nabla_\mu R^\star$.

Some examples are studied by Coley and McManus \cite{McManus94}, with the energy-momentum tensor \eqref{EMT} with $q_i=0$.
They show that if $u_i$ is shear-free, vorticity-free and geodesic, then the spacetime is GRW.
%The case $\varphi =0$ specialises a DW to a static spacetime. 

\section{Spherical doubly warped spacetimes}
From now on we restrict to DW spacetimes with spherical symmetry:
\begin{align}
 ds^2 = -b^2(r) dt^2 + a^2(t) \left [f_1^2(r)dr^2 + f_2^2(r) d^2\Omega^2_{n-2} \right ] \label{F1F2}
 \end{align} 
The coordinates of $(M^\star , g^\star)$ are $r$ and the $n-2$ 
angles $\theta_a$ of the sphere $S_{n-2}$. The space metric tensor is diagonal with $g^\star_{rr}=f_1^2(r)$ 
and $g^\star_{aa} = f_2^2(r) g^2_a (\boldsymbol\theta)$.\\
The main results that are obtained: the conformal flatness of $M^\star$,  the covariant forms of the electric and Ricci tensors.

 \begin{prop} $M^\star$ is conformally flat $(C^\star_{\mu\nu\rho\sigma}=0)$.
\begin{proof}
The change of variable $d\rho =f_1(r)dr$, makes the metric $g^\star $ warped: 
$d s_\star^2=d\rho^2 + (f_2\circ r)(\rho)^2 d\Omega_{n-2}^2$.
The fiber $S_{n-2}$ is a constant curvature submanifold, then  $M^\star$ is conformally flat by Theorem 1.i (warped manifolds) in ref.\cite{BV} by 
Brozos-Vazquez  et al. 
\end{proof}
\end{prop}
In spherical coordinates, the unit radial vector on $M^\star $ is
\begin{align*}
 n_r = f_1(r), \qquad n_a=0 \quad (a=1,...,n-2). % \label{ennesphere}.
\end{align*}
\begin{prop}\label{nablanf1f2}
The unit radial vector is torse-forming:
\begin{align}
&\nabla^\star_\mu n_\nu = \frac{\Theta}{n-2}(g^\star_{\mu\nu} - n_\mu n_\nu) \label{torsen1}\\
&\Theta= \frac{n-2}{f_1} \frac{d}{dr}\log f_2 \label{torsen2}
\end{align}
$\partial_\mu \Theta $ is proportional to $n_\mu $.
\begin{proof}
The equations are checked in spherical components with the Christoffel symbols in Appendix \ref{SPHEREMETRIC}:

$\nabla^\star_r n_r = \partial_r f_1 - \Gamma_{rr}^{\star r} f_1 =0 = \partial_r f_1 - 
\tfrac{f_1}{2} g^{\star rr} \partial_r g^\star _{rr} =0$,

$\nabla^\star_r n_a = - \Gamma_{r a}^{\star r} f_1= - \tfrac{1}{2}g^{\star rr} \partial_a g^\star_{rr}=0$.

$\nabla^\star_a n_{a'} = -\Gamma_{a a'}^{\star r} f_1= \delta_{aa'} g^2_a ({\boldsymbol\theta})\frac{f_2}{f_1}\frac{df_2}{dr} $\\
In the last equation one reads $\Theta = (n-2) (1/f_1) \partial_r \log f_2$. 
Since it is a function only of $r$, it is
$\partial_\mu \Theta =( n_\mu /f_1)(\partial_r \Theta)$. 
\end{proof}
\end{prop}
In the coordinates \eqref{sphere} the warping function $b$ only depends on $r$. If it is not a constant, then:
$\dot u_0=0$ because $u^k\dot u_k=0$, and
$$  \dot u_\mu = u^0\nabla_0 u_\mu = -u^0\Gamma_{0,\mu}^0 u_0 = \delta_{\mu r} \frac{b_r}{b} $$ 
i.e. $\dot u_\mu$ is a radial vector, with components $\dot u_r = b_r/b$, $\dot u_a=0 $, $a=1\ldots n-2$.  
The parameter $\eta =\dot u_i \dot u^i = g^{rr} (\dot u_r)^2$ is a factored scalar function of $t$ and $r$:
\begin{align}
\eta = \frac{1}{a^2(t)} \frac{b_r^2}{f_1(r)^2 b(r)^2}. \label{eta}
\end{align}
The vector field ${\hat n}_\mu = \dot u_\mu/(a\sqrt \eta)$ in eq.\eqref{enne} is  radial and coincides with $n_\mu$.
The equations \eqref{torsen1} and \eqref{torsen2} for $n_\mu $ are compared with eq.\eqref{nablaenne} for ${\hat n}_\mu$. They show that,
by the spherical symmetry,
$$w_\mu=0, \qquad \Pi_{\mu\nu}=0, \qquad  \partial_\mu \Theta \propto n_\mu. $$
Therefore $(M^\star, g^\star)$ is a warped manifold (the third class in Prop.~\ref{nablaenne}).

 The integrability condition in $M^\star$ and the property $C^\star_{\mu\nu\rho\sigma}=0$ 
 yield the Ricci tensor $R^\star_{\mu\nu} $ and the eigenvalue $\xi^\star$ (see Appendix\ref{INTEGRABILITY}):
\begin{align}
R^\star_{\mu\nu}=& \xi^\star n_\mu n_\nu+\frac{R^\star-\xi^\star}{n-2} (g^\star_{\mu\nu}-n_\mu n_\nu )\label{RicciqE} \\
\xi^\star =& 
%& -\frac{\Theta^2}{n-2}- n^\mu\partial_\mu \Theta \\&  
-\frac{n-2}{f_1^2}[ (\partial_r \log f_2)^2 +  \partial^2_r \log f_2 -
(\partial_r \log f_1)(\partial_r \log f_2)]   \label{xistar}
\end{align}
%The Ricci tensor of $M^\star $ is a quasi-Einstein, 
%with $R^\star = g^{\star \mu\nu}R^\star_{\mu\nu}$ and $R^\star_{\mu\nu} n^\nu =\xi^\star n_\mu$. \\

\subsection{The electric tensor} Although $M^\star$ is conformally flat, the whole spacetime in general is not. The electric tensor arises solely from the Weyl tensor of the spacetime.
\begin{prop}
\begin{gather}
E_{\mu\nu} = E(r) \left [n_\mu n_\nu - \frac{g^\star_{\mu\nu}}{n-1} \right ] \label{Emunu}\\
E(r)=\frac{n-3}{n-2}\frac{1}{f_1^2}\left[ \frac{f_1^2}{f_2^2} + 
\frac{d^2}{dr^2} \log f_2 - \frac{f_1'f_2'}{f_1f_2}  +\frac{b'}{b}\frac{d}{dr} \log (f_1f_2) 
 - \frac{b''}{b}  \right ]  \label{E(r)}
 \end{gather}
 where a prime is a derivative in $r$.
\begin{proof}
The electric tensor is spacelike and can be obtained from the general relation 
\eqref{GebaExt}. To evaluate $\nabla^\star_\mu b_\nu$ note that $b_\nu = n_\nu  b_r/f_1$. Then:
\begin{align*}
 \nabla^\star_\mu b_\nu &= n_\nu \partial_\mu \frac{b'}{f_1} + \frac{b'}{f_1}\nabla^\star_\mu n_\nu
 = n_\nu n_\mu \frac{1}{f_1}\frac{d}{dr} \frac{b'}{f_1} + \frac{b'}{f_1} \frac{\Theta}{n-2}(g^\star_{\mu\nu} - n_\mu n_\nu)\\
& =n_\nu n_\mu \frac{1}{f_1^2}\left [b'' -   b' \frac{d}{dr} \log (f_1 f_2)
\right ] + g^\star_{\mu\nu}  \frac{1}{f_1^2} b' \frac{d}{dr} \log f_2 
\end{align*}
In particular: 
\begin{align}
 \nabla^\star_\mu b^\mu  = \frac{1}{f_1^2} \left [b'' -   b'\frac{d}{dr}\log (f_1 f_2)
 + (n-1)  b' \frac{d}{dr} \log f_2\right  ] \label{divdotu}
 \end{align} 
 Then:
\begin{align}
\frac{1}{b}\left[\nabla^\star_\mu b_\nu - g^\star_{\mu\nu} \frac{\nabla^\star_\sigma b^\sigma}{n-1}\right ] = 
\left[ n_\nu n_\mu - \frac{g^\star_{\mu\nu}}{n-1} \right ] \frac{1}{f_1^2}\left [\frac{b''}{b} -   \frac{b'}{b} \frac{d}{dr} \log (f_1 f_2)
\right ] 
\end{align}
Since $R^\star_{\mu\nu}$ in \eqref{RicciqE} is quasi-Einstein and $\nabla^\star_\mu b_\nu$ has the same tensor form, the space components of
the electric tensor are $E_{\mu\nu}(r)=E_1 g^\star_{\mu\nu}+ E_2 n_\mu n_\nu$ (see eq.\eqref{GebaExt}). Being traceless it is 
$0=(n-1)E_1+E_2$ so that $E_{\mu\nu}$ has the form \eqref{Emunu}.\\
The contraction of \eqref{GebaExt} with $n^\mu$ gives:
\begin{align*}
\xi^\star - \frac{R^\star}{n-1} + \frac{(n-2)(n-3)}{n-1} \frac{1}{f_1^2} 
\left[\frac{b''}{b} - \frac{b'}{b} \frac{d}{dr} \log (f_1f_2) \right ] = -\frac{(n-2)^2}{n-1} E(r),
\end{align*}
where $R^\star $ is evaluated in Appendix \ref{SPHEREMETRIC} ($N=n-2$). This and eq.\eqref{RicciqE} for $\xi^\star$ give:
\begin{align}
\xi^\star-\frac{R^\star}{n-1} = 
-\frac{(n-2)(n-3)}{(n-1)}\left[ \frac{1}{f_2^2} + 
\frac{1}{f_1^ 2}\frac{d^2}{dr^2} \log f_2 - \frac{1}{f_1^2}\frac{f'_1f_2'}{f_1f_2}    \right ]. \label{Rxistar}
\end{align}
Insertion in the previous equation completes the evaluation of $E(r)$.
\end{proof}
\end{prop}

Since $E_{ij}u^i=0$, $E^k{}_k=0$, the extension to spacetime of the expression \eqref{Emunu} is
\begin{align}
E_{ij} = \frac{E(r)}{a^2(t)} \left [N_i  N_j - \frac{g_{ij}+ u_iu_j}{n-1} \right ] \label{ECOV}
\end{align}
where in spherical coordinates $N_i=(0, a n_\mu)$, $N_iN_j g^{ij}=a^2n_\mu n_\nu g^{\mu\nu}=1$, $n_\mu n_\nu$ is the radial projector 
in $M^\star $. It is 
$$ E_{ij} E^{ij} =  \frac{n-2}{n-1}\frac{E (r)^2}{a^4(t)}. $$
If $\dot u_i $ exists ($b(r)$ is non-constant), then
$N_i N_j =\frac{\dot u_i\dot u_j}{\eta} $ and  
$\dot u$ is an eigenvector of the electric tensor: $E_{ij}\dot u^i = \frac{n-2}{n-1}\frac{E(r)}{a^2(t)} \dot u_j$.

\subsection{The Ricci tensor}
The covariant expression \eqref{ECOV} is inserted in \eqref{RicciRicci} to obtain the general expression of the Ricci tensor for the metric \eqref{F1F2}.\\
$\bullet $ If $\dot u_i\neq 0$: 
\begin{align}
R_{jk} =& u_j u_k (-\xi+\nabla_p\dot u^p)
+ \frac{h_{jk}}{n-1} (R-\xi  + \nabla_p \dot u^p) - (n-2)\varphi (u_j \dot u_k + \dot u_j u_k)
\nonumber\\
& +\left [\frac{\dot u_j \dot u_k}{\eta} -\frac{h_{jk}}{n-1}\right ] \left [\nabla_p \dot u^p -(n-1) \left ( \eta + \frac{\dot u^i\nabla_i\eta}{2\eta} \right ) - (n-2) \frac{E(r)}{a^2(t)}  \right ] \label{RicciSphere}
\end{align} 
where $h_{jk}=g_{jk}+u_j u_k$.
It only involves the vectors $u_i$ and $\dot u_i$, and scalar fields. \\
It is the sum of a quasi-Einstein (perfect fluid) term, a current term, and a traceless anisotropic term that comes form two sources: the acceleration and the electric tensor. This anisotropy occurs despite the metric being spherically
symmetric.\\
%In general, it has three distinct eigenvalues.
%This one has degeneracy $n-2$: 
%$$ \lambda_3 = \frac{R-n\xi}{n-1} +\eta + \frac{\dot u^i\nabla_i\eta}{2\eta} + \frac{n-2}{n-1} \frac{E(r)}{a^2(t)} $$
%with spacelike eigenvectors orthogonal to $u$ and $\dot u$. $\lambda_1$ and $\lambda_2$ are real if they 
%correspond to a timelike and a spacelike eigenvector, otherwise they are complex conjugate.\\
%
$\bullet$ If $\dot u_i=0$, eq.\eqref{RicciSphere} is the Ricci tensor on a spherical GRW spacetime:
\begin{align}
R_{jk} = -\xi u_j u_k + \frac{R-\xi}{n-1} h_{jk}
 -(n-2) \left [N_jN_k -\frac{h_{jk}}{n-1}\right ]  \frac{E(r)}{a^2(t)} . \label{RicciSphereGRW}
\end{align} 
The eigenvalues are: $\xi$, $\frac{R-\xi}{n-1}- \frac{(n-2)^2}{n-1} \frac{E(r)}{a^2}$ and $\frac{R-\xi}{n-1}+ \frac{n-2}{n-1} \frac{E(r)}{a^2}$, with degeneracies 
$1,1,n-2$. The anisotropy is totally due to the electric term. $R_{jk}$  has the perfect fluid form if and only if $E(r)=0$.

\begin{lem}[this formula is proven in Appendix \ref{APPLEMMAACC}]\label{LEMMAACC}
\begin{align}
\nabla_p \dot u^p - (n-1)\left( \frac{\dot u^i\nabla_i \eta}{2\eta}  + \eta \right )  = 
- \frac{n-2}{a^2(t) f_1^2 b}\,\left [b''  - b' \frac{d}{dr} \log (f_1f_2)  \right ].
 \end{align} 
 \end{lem}
Other useful expressions in the spherical comoving coordinates are collected below. $R$ and $R^\star$ are \eqref{RRstar}  and \eqref{RstarStar} with $N=n-2$ (number of angles). $\nabla_p\dot u^p$ is \eqref{nablapdotup}, with $\nabla^\star_\mu b^\mu$ in \eqref{divdotu}:
\begin{align}
%&\varphi = \frac{\nabla_j u^j}{n-1} = \frac{1}{n-1} \Gamma_{\mu 0}^\mu u^0 =  \frac{a_t}{ab},\\
&\dot\varphi = u^0\partial_t \varphi =\frac{1}{b}\partial_t  \frac{a_t}{ab} = \frac{a_{tt}}{ab^2} - \frac{(a_t)^2}{a^2b^2}, \quad \xi \equiv (n-1)(\varphi^2+\dot \varphi)=(n-1)\frac{1}{b^2} \frac{a_{tt}}{a}, \\
&\nabla_p \dot u^p = \frac{1}{a^2b}\nabla^\star_\nu b^\nu = \frac{1}{a^2b}\frac{1}{f_1^2}\left [  b'' - b' \frac{d}{dr}
\log (f_1f_2) + (n-1) b' \frac{d}{d r}\log f_2\right ],\\
&R=\frac{R^\star}{a^2}  - \frac{2}{a^2b}\nabla^\star_\nu b^\nu + \frac{n-1}{a^2 b^2}\left [(n-2) a_t^2  +2 a a_{tt}\right].
\label{CurvatureScalar}
%&\nabla^\star_\nu b^\nu =\frac{1}{f_1^2}\left [ \frac{\partial^2 b}{\partial r^2} - \frac{\partial b}{\partial r}\frac{\partial}{\partial r}\log (f_1f_2) + (n-1) \frac{\partial b}{\partial r}\frac{\partial}{\partial r}\log f_2\right ]
\end{align}

\section{Radiating stars}
The Ricci tensor \eqref{RicciSphere} and the Einstein equations $R_{ij} - \tfrac{1}{2}R g_{ij}=T_{ij}$ fix the structure of the energy-momentum tensor of  matter and radiation:
\begin{align}
& T_{ij}=  \mu u_i u_j + P h_{ij} + (u_iq_j + u_j q_i ) + \left [ N_i N_j-\frac{h_{ij}}{n-1}\right ] (p_r -p_\perp) \label{Tij}\\
&\mu=-\xi+\nabla_p\dot u^p +\tfrac{1}{2}R \\
&(n-1)P= -\xi +\nabla_p \dot u^p - \tfrac{1}{2}(n-3)R \\
&q_j=-(n-2)\varphi \dot u_j \\
&p_r-p_\perp = \nabla_p \dot u^p - (n-1) \left ( \eta + \frac{\dot u^i\nabla_i\eta}{2\eta} \right )- (n-2) \frac{E(r)}{a^2(t)}  
\end{align}
$T_{ij}$ describes an anisotropic fluid with energy density $\mu$, heat flow $q_j$,  shear-free and vorticity-free velocity $u_j$, radial pressure $p_r$, tangential pressure $p_\perp $, effective pressure 
$P= \frac{1}{n-1}p_r + \frac{n-2}{n-1} p_\perp $.

For a radiating star the spacetime splits into an interior spherical DW manifold $V^-$ describing the fluid history, 
and an exterior manifold $V^+$ described by the spherically symmetric Vaidya solution \eqref{Vaidya1}
of the Einstein equations with a pure radiation energy-momentum tensor. A change of coordinates
%$ v = t + r + 2m\log (\frac{r}{2m} -1) $
gives the metric the convenient form (in dimension $n\ge 4$ \cite{Bhui}):
\begin{align}
ds^2_+ = -\left(1-\frac{2m(v)}{(n-3){\sf R}^{n-3}}\right ) dv^2 -2 dv d{\sf R} + {\sf R}^2 d\Omega^2_{n-2} \label{Vaidya}
\end{align}
where $m(v)$ is the Vaidya exterior mass. %$k_i = u_i - \dot u_i/\sqrt{\eta}$. 
%Then:$$ T_{ij}^{\rm ext} = T \left [u_i u_j - \frac{1}{\sqrt \eta} u_i \dot u_j  - \frac{1}{\sqrt \eta} u_j \dot u_i + \frac{1}{\eta} 
%\dot u_i \dot u_j \right ]$$
%
\subsection{Boundary conditions} 
The inner solution \eqref{F1F2} and Vaidya solution match at the boundary $\Sigma $ of the collapsing star. 
It is a time-like hypersurface with intrinsic metric
$$ ds^2  = -d\tau^2 + {\mathscr R}^2(\tau) d\Omega^2_{n-2} $$
In the two coordinates, the boundary has equations 
$f^- (t,r,\boldsymbol \theta) = r-r_\Sigma=0$ where $r_\Sigma $ is {\em constant}, 
and $f^+(v,{\sf R},\boldsymbol\theta)= {\sf R}-{\sf R}_\Sigma(v)=0$. The metric and its first derivatives are continuous 
throughout the spacetime. Then $g_{ij}^\pm$ and the extrinsic curvature tensors in $V^\pm$ have to match at $\Sigma $. 
We proceed with the prescriptions well described by Santos in ref.\cite{Santos85} with the generalization $n\ge 4$ in \cite{Bhui}. 

The continuity of the metric at $\Sigma $ is expressed by
\begin{align}
{\mathscr R}^2(\tau) = a^2(t) f_2^2(r_\Sigma)= {\sf R}_\Sigma^2(v) \label{CONTIN}
\end{align}
$$d \tau^2 = b^2(r_\Sigma) dt^2 = \left (1-\frac{2m}{(n-3){\sf R}_\Sigma^{n-3}} +2  \frac{d{\sf R}_\Sigma}{dv}\right ) dv^2 $$
The latter gives:
\begin{align}
\frac{dt}{d\tau} = \frac{1}{b(r_\Sigma)}, \qquad \left (\frac{dv}{d\tau}\right )^{-2} = 1-\frac{2m}{(n-3){\sf R}_\Sigma^{n-3}} +2 \frac{d{\sf R}_\Sigma}{dv}. \label{57}
\end{align}
%The gradient vectors
%$(0,1,{\bf 0}_{n-2})$ and $(-\frac{d{\sf R}}{dv},1,{\bf 0}_{n-2})$ of $f^\pm$ are orthogonal to $\Sigma $. They are %normalized in the respective metrics: 
%$$ n^- = (0, a(t)f_1(r_\Sigma), {\bf 0}), \qquad n^+ = (0, ??) $$
The relevant second forms for the interior solution are evaluated, while those for the Vaidya solution are given in \cite{Bhui}. 
The continuity condition $K_{\theta\theta}^- = K_{\theta\theta}^+$ is:
\begin{align}
 a(t) \frac{f_2}{f_1} \frac{d f_2}{dr}\Big |_{r=r_\Sigma}  =& \frac{dv}{d\tau} {\sf R}_\Sigma \left[ 1-\frac{2m}{(n-3){\sf R}_\Sigma^{n-3}}\right ]
+ {\sf  R}_\Sigma \frac{d {\sf R}_\Sigma}{d\tau}\nonumber
\end{align}
Now use ${\sf R}_\Sigma =a f_2$: $\frac{d{\sf R}_\Sigma}{d\tau} =f_2 \frac{da}{d\tau}= f_2(r_\Sigma)a_t /b$. 
The previous equation becomes:
\begin{align}
 \frac{1}{f_1} \frac{d f_2}{dr} - \frac{a_t}{b} f_2
  = \frac{dv}{d\tau}  \left[ 1-\frac{2m}{(n-3){\sf R}_\Sigma^{n-3}}\right ] \label{58}
  \end{align}
Multiply the second equation in \eqref{57} by $\frac{dv}{d\tau}$:  $\frac{d\tau}{dv} = [1-\frac{2m}{(n-3){\sf R}_\Sigma^{n-3}}]
\frac{dv}{d\tau}  +2 \frac{d{\sf R}_\Sigma}{d\tau}$ and simplify the above equation:
\begin{align*}
 \frac{1}{f_1} \frac{d f_2}{dr} + \frac{a_t}{b} f_2
  = \frac{d\tau}{dv} ; 
  \end{align*}
 With eq.\eqref{58} the Misner-Sharp mass function is obtained in terms of the inner solution:
 \begin{align}
m(t,r)=\frac{n-3}{2}(a f_2)^{n-3}\frac{f_2^2}{f_1^2}\left[\frac{f_1^2}{f_2^2}- \left( \frac{f_2'}{f_2}\right)^2 + f_1^2\frac{a_t^2}{b^2}  \right]_\Sigma \label{59}
  \end{align}
The continuity condition $K_{\tau\tau}^- = K_{\tau\tau}^+$ is:
\begin{align*}
-\frac{1}{f_1 a} \frac{b'}{b}\Big |_\Sigma = \frac{d^2v}{d\tau^2} \left(\frac{dv}{d\tau}\right)^{-1}
 - \frac{dv}{d\tau} \frac{m}{{\sf R}_\Sigma^{n-2}}
 \end{align*}
 This gives, up to an common prefactor $1/(a^2b)$, the equation 
% \begin{align}
%2\frac{1}{f_2} \frac{d f_2}{dr}  \frac{1}{b} \frac{db}{dr}+ 2 a_t  \frac{f_1}{b^2} \frac{db}{dr} - 2\frac{f_1^2}{b^2} a a_{tt} 
%= (n-3)  \left[\frac{f_1^2}{f_2^2}- \frac{1}{f_2^2} \left(\frac{d f_2}{dr}\right)^2 +f_1^2\frac{a_t^2}{b^2}  \right]  
 %\end{align}  
 \begin{align}
2 a a_{tt} +(n-3)a_t^2- 2 a_t  \frac{b'}{f_1} 
+ (n-3) \frac{b^2}{f_1^2} \left[\frac{f_1^2}{f_2^2}- \left(\frac{f_2'}{f_2}\right)^2\right ] 
-2\frac{bb'}{f_1^2} \frac{f_2'}{f_2}  =0.  \label{BEQ}
 \end{align} 
 Since the functions are evaluated at $r_\Sigma $, it is a differential equation for $a(t)$ with constant coefficients: $2 a a_{tt} +(n-3)a_t^2  + A a_t - B =0 $.\\
 A simple solution is $a(t)=-Ct$. The non linear equation was studied by Paliathanasis et al. \cite{Paliathanasis}. They obtained the expansion near a movable singularity (i.e. dependent on initial conditions) that, extended to $n\ge 4$ is: $a(t) = a_0 (t-t_0)^\nu + a_1(t-t_0) + a_2 (t-t_0)^{2-\nu} + ... $ 
 with exponent $\nu = \frac{2}{n-1} $, arbitrary $a_0$, $a_1=- \frac{1}{2} \frac{A(n-1)}{(n-2)(n-3)}$ etc.
 An integral is obtained with $a_t=y(x)$,  $x=e^a$, with equation $ 2\frac{dy}{dx} +(n-3)y + A - B/y=0 $. 
Eq.\eqref{BEQ} translates to a relation for the physical parameters at the surface of the star. 
A rather long evaluation shows that \eqref{BEQ} is equivalent to 
\begin{align}
 (p_r - \sqrt{q_jq^j} )_\Sigma =0
 \end{align}
At the boundary, the radial pressure equals the heat current intensity.

\section{Pressure isotropy} 
Several models for radiating stars are characterized by isotropic pressure, $p_r=p_\perp$, i.e.
a zero anisotropic term in the Ricci tensor. \\
With Lemma \ref{LEMMAACC} and eq.\eqref{Emunu} for $E(r)$, the condition for isotropy is the linear
differential equation for the function $b(r)$:
 \begin{align}
 0= \frac{d^2 b}{dr^2} -\frac{d b}{dr} \frac{d}{dr}\log (f_1f_2) +(n-3)b \left [\frac{f_1^2}{f_2^2}
- \frac{f_1'f_2'}{f_1f_2}  +\frac{d^2}{dr^2}\log f_2 \right ] \label{VANISHINGTOTAL}
\end{align}
Solutions are obtained by setting to zero the acceleration term and the electric term of anisotropy separately, or as a whole.
The following propositions give conditions for the vanishing of single terms. %
\begin{prop}[Condition I] \label{PP0} If $\dot u_i\neq 0$:
\begin{align}
\frac{d^2 b}{dr^2} -\frac{d b}{dr} \frac{d}{dr}\log (f_1f_2) =0 \; \Longleftrightarrow \;
 \frac{d b(r)}{dr} = 2K  f_1(r)f_2 (r) 
 \end{align} 
with constant $K$.
\end{prop}

\begin{prop}[Condition II]\label{PP2}
\begin{align}
 \frac{f_1^2}{f_2^2} - \frac{f'_1 f_2'}{f_1f_2}  + \frac{d^2}{dr^2} \log f_2=0 \; &\Longleftrightarrow \;
 f_1^2(r) = \frac{f'_2(r)^2}{1+ C f_2(r)^2} \label{44}
 \end{align} 
with a constant $C$ ensuring $f_1^2>0$. 
\begin{proof}
Eq.\eqref{44} is a Bernoulli equation of the form $f'_1 + A(r) f_1 +B(r)f_1^3 =0$.
Divide by $f_1^3$ and set $y=1/f_1^2$. The equation now is: $ y' - 2y A - 2B =0$. 
\end{proof}
\end{prop}
Together, conditions I and II ensure isotropy. After fixing $f_2$, one evaluates $f_1$ with Condition II and then 
$b$ with Condition I.

\begin{remark}\quad\\
$\bullet$ If $\dot u_i=0$ then Condition II is necessary and sufficient for isotropy.\\
$\bullet$ If $\nabla_m C_{jkl}{}^m=0$ or $C_{jklm}=0$ then $E_{ij}=0$ and {\rm Condition I} assures isotropy.\\
$\bullet$ The Ricci tensor of $M^\star $ has eigenvalues $\xi^\star $ and $(R^\star -\xi^\star)/(n-2)$ with degeneracy $n-2$. When they are equal, $\xi^\star = R^\star/(n-1)$, then  $R_{\mu\nu}^\star $ is Einstein. 
With eq.\eqref{Rxistar} one obtains the equivalence: 
\begin{align}
R^\star_{\mu\nu} = \frac{R^\star}{n-1}g^\star_{\mu\nu} \;\Longleftrightarrow \; 
\text{\rm Condition II}
\end{align}
$\bullet$ By Lemma  \ref{LEMMAACC}, {\rm Condition I} is equivalent to $0=\nabla_p \dot u^p - (n-1)\left( \frac{\dot u^i\nabla_i \eta}{2\eta}  + \eta \right ) $. 
\end{remark}

\begin{prop}\label{CORSPECIAL}
If $f_2=rf_1$ and if {\rm Condition II} is true, then
\begin{align}
 f_1(r) =  \frac{K}{r^2 + C} 
 \end{align}
 If also {\rm Condition I} is true then $b(r)\propto f_1(r)$
 %$$f_1(r) = \frac{K}{r^2+C}, \qquad b(r) = \frac{1}{2}\frac{\tilde K K^2}{r^2 + C}  $$
and the metric is conformally Robertson-Walker, with flat $M^\star $:
$$ ds^2 = \frac{K^2}{(r^2 + C)^2} \left[ -dt^2 + a^2(t)(dr^2 + r^2 d^2\Omega_{n-2} )\right ] $$ 
\begin{proof}
The equation in Prop.\ref{PP2} now is:
$ \frac{1}{r^2} + \partial^2_r\log(r f_1) - (\partial_r\log f_1) \partial_r \log (rf_1)=0$.
With $Y=\partial_r \log f_1$ the equation becomes: $Y' - \frac{1}{r}Y - Y^2 =0$. Divide by $Y^2$, multiply by $r$ and put
$X=1/Y$: $(rX)'+r=0$. Integration gives $rX(r)+r^2/2 + C/2=0$. Going back to $Y$ we obtain
$$  \frac{d}{dr}\log f_1 =-\frac{2r}{r^2 + C} = -\frac{d}{dr}\log (r^2 + C) $$
 Another integration yields the result.
\end{proof}
\end{prop}

The following solution of eq.\eqref{VANISHINGTOTAL} for isotropy was obtained by
Wagh et al. \cite{Wagh01}. It is parameterized by $f_2(r)$ with $f_2'>0$, and is checked by substitution. 
The extension to $n\ge 4$ is  
\begin{prop}\label{WAG}
\begin{align}
b(r) = K f_2(r), \qquad  f_1(r) = \sqrt{\frac{n-2}{n-3}}\; \frac{df_2}{dr}, 
\end{align}
$K>0$ is a constant. If also $f_2=rf_1$ then $f_2(r) = C r^\nu$, $\nu = \sqrt{\frac{n-3}{n-2}}$.
\end{prop}

Eq.\eqref{VANISHINGTOTAL} simplifies with $f_2 (r)= rf_1(r)$, that applies to most of the physics models 
considered here:
\begin{align}
0=\frac{d^2 b}{dr^2} -\frac{d b}{dr} \frac{d}{dr}\log (r f_1^2) +(n-3)b \left[\frac{d^2\log f_1}{dr^2} - \frac{1}{r} \frac{d \log f_1}{dr} - \left ( \frac{d\log f_1}{dr}\right )^2 \right ]  \label{VANISHINGTOTAL2}
\end{align}
The change $x=r^2$ and $F=1/f_1(x)$ transforms it into the isotropy condition found by Glass \cite{Glass} in $n=4$, and by Banerjee \cite{Banerjee04} in higher dimensions:
\begin{align}
0= \frac{d^2 b}{dx^2} F + 2 \frac{db}{dx}\frac{dF}{dx} -(n-3) b \frac{d^2F}{dx^2} 
\label{EQBANERJEE}
 \end{align}
With an input function, the differential equation is solved for the other function. Simple pairs are found 
with $F_x=0$, $F_{xx}=0$ and $b_{xx}=0$ and are the first three lines of Table I, with other solutions in the literature.
\begin{table}[h!]
\centering
\begin{tabular}{| c | c | c | l |} 
%\hline
$f_1(r)$ &  $b(r)$ & Notes  &ref.  \\
\hline
 $1$     &    $ z $  & $ E(r)=0$              & \cite{Modak84}\cite{Banerjee02}\cite{Banerjee04}\\
$ z^{-1} $ &  $a+b z^{-1} $ & $E(r)= 0$ &\cite{Maiti}\cite{Banerjee89}\cite{DeOliveira85}\cite{Schafer00}\\
%{} &$ (1+kr^2)^{-1} $ & $ 0 $ & $ \alpha-\beta \frac{1-r^2}{1+r^2} $ & De Oliveira \cite{DeOliveira85}\\
%\hline
 $z^{-\nu}$     &   $ z$  & $\nu =\frac{n-1}{n-3}$              &  \cite{Banerjee04}\\
 $(a\sqrt z+1)^2 /(bz) $ & $(a\sqrt z -1)/(a\sqrt z +1)$ & $n=4$ &\cite{Banerjee90} \\
% $z^{\mu-\frac{1}{2}}$     &    $ b_1z^{q_1}+b_2 z^{q_2} $  & $q^2-2q\mu+\frac{n-3}{4}=0 $   & Tewari ?\cite{Tewari13}\\
  $z^{-q-\frac{1}{2}}$     &    $ z^p$  & $p^2+2qp=(n-3)(q^2-\tfrac{1}{4}) $  
  & \cite{Tewari13}\\
\hline
\end{tabular}
\caption{Isotropic solutions with $f_2(r)=r f_1(r)$. $z\equiv 1+Kr^2$, where $K$ is a constant.
The solutions with $E(r)=0$ satisfy conditions I and II.
Constants may change by rescaling  $t$ and $r$.}
\end{table}

\noindent
Here are other pairs, with $f_2=rf_1$:
\begin{align}
&f_1(r)=  \frac{K_1}{K_2 + (b_0+b_2 r^2)^\nu},\quad  b(r)=b_0+b_2 r^2,  \quad \nu=\frac{n-1}{n-3} \\
&f_1(r) = K r^{p-1}, \quad b(r) = K_1 r^{q_1}+ K_2 r^{q_2},\quad
 q_{1,2}=p\pm \sqrt{p^2(n-2)-(n-3)} \\
 &f_1(r) = e^{- K r^2}, \quad  b(r) = e^{- K (1\pm \sqrt{n-2}) r^2}
\end{align} 
The last pair belongs to a larger new family discussed in Appendix \ref{SOLLM}. 
Oscillating and hyperbolic solutions are found in \cite{Sanyal84}. Other solutions
are studied in \cite{Tewari15}.

\begin{example}[Banerjee and Chatterjee \cite{Banerjee04}] The model describes the gravitational
collapse of a star, started at $t=-\infty$.
$$ ds^2=-(1+Kr^2)^2 dt^2 + C^2t^2 (dr^2 + r^2 d\Omega^2_{n-2}) $$
The subspace $M^\star $ is flat, the spacetime $M$ is conformally flat ($E(r)=0$). 
Condition I is satisfied: the model is isotropic. The particular solution $a(t)=-Ct$ is chosen, then $\xi =0$. 
%We evaluate: 
%$\nabla_\nu^\star b^\nu = 2K(n-1)$ Then $\nabla_p \dot u^p = \frac{2K (n-1)}{a^2 (1+Kr^2)}$
The scalar curvature, the energy density and the isotropic pressure are:
$$ R= - \frac{4K (n-1)}{C^2t^2 (1+Kr^2)} + \frac{(n-1)(n-2)}{(1+Kr^2)^2t^2} $$
$$\mu = \frac{(n-1)(n-2)}{2t^2(1+Kr^2)^2} , \quad p= \frac{2K(n-2)}{C^2t^2 (1+Kr^2)} -\frac{1}{2}\frac{(n-2)(n-3)}{(1+Kr^2)^2t^2} $$
\end{example}

\begin{example}[Wagh et al. \cite{Wagh01}]\label{EXWAGH} The model describes the collapse of a radiating star with equation of state
$p=w\mu$. The isotropy condition is solved with the metric in Prop.\ref{WAG}. We generalize their results to $n\ge 4$:
\begin{align}
 ds^2 = -K^2f_2^2(r)dt^2 + a^2(t) \left [ \frac{n-2}{n-3} \left (\frac{df_2}{dr}\right )^2 dr^2 + f_2(r)^2 d\Omega_{n-2}^2 
\right ]\label{Waghmetric}
\end{align}
The following geometric quantities are evaluated, and depend on $a(t)$ and $f_2(r)$:
$$R^\star = \frac{n-3}{f_2^2} , \qquad \xi = \frac{n-1}{K^2 f_2^2} \frac{a_{tt}}{a}, 
%\qquad \nabla^\star_\nu b^\nu =K \frac{n-3}{f_2(r)} $$ 
%\nabla_p \dot u^p = \frac{n-3}{f_2^2(r) a^2(t)}\quad 
 \qquad E(r) = \left(\frac{n-3}{n-2} \right )^2 \frac{1}{f_2^2} $$
\begin{align}
 R= -\frac{n-3}{a^2 f_2^2} +\frac{n-1}{K^2 a^2 f_2^2} \left [(n-2) a_t^2 +2a a_{tt} \right ] \label{RWagh}
 \end{align}
Being $E\neq 0$, this model is never conformally flat. In the Einstein equations, the fluid parameters are
$$ \mu = \frac{n-3}{2a^2 f_2^2} +\frac{(n-1)(n-2)}{2K^2 f_2^2} \frac{a_t^2}{a^2}, 
\quad p= \frac{n-3}{2a^2 f_2^2} - \frac{n-2}{2K^2 a^2 f_2^2} \left[ (n-3)a_t^2 + 2aa_{tt}\right ]$$
$$q_r = -\frac{n-2}{f_2^2}\frac{df_2}{dr}\frac{a_t}{a}$$
Note that $\mu>0$ and $\frac{d}{dr}\mu = -2\mu \frac{d}{dr}\log f_2$. The Misner-Sharp mass is
$$ m(t) = \frac{1}{2}(n-3) (a f_2(r_\Sigma))^{n-3} \left [ \frac{1}{n-2}- \frac{a_t^2}{K^2} \right ] $$   
The equation of state $p=w\mu$
and the junction condition \eqref{BEQ} are two equations for $a(t)$ and $w$:
\begin{align}
&2a a_{tt} +  [(n-1)w+(n-3)]a_t^2 +\frac{n-3}{n-2}(w-1) K^2=0 \nonumber\\
&2 aa_{tt} +(n-3)a_t^2 - 2K a_t\sqrt{\frac{n-3}{n-2}}-\frac{n-3}{n-2} K^2 =0 \label{Waghboundary}
 \end{align}
The solutions are: $ w^2 (n^2-3n+3)-2w-(n-3)=0$, and $a(t)$
with $a_t<0$ for a collapsing star: 
$$ a_t = -K  \sqrt {\frac{n-3}{n-2} } \sqrt{\frac{1-w}{(n-3)+(n-1)w}}$$ 
The linear $a(t)$ means $\xi =0$.
\end{example}

\begin{example}[Euclidean DW stars]\label{EUCL}
The Euclidean metric $ds^2 = -B^2(t,r) dt^2 + R'^2(t,r)dr^2 + R^2(t,r) d\Omega^2_{n-2}$ was introduced by Herrera and Santos \cite{Herrera10}. 
In particular, a Euclidean DW spherical metric has the form  
$$ ds^2 = -b^2 (r) dt^2 + a^2(t)[f'^2(r) dr^2 + f^2 (r) d\Omega^2_{n-2}]$$
The surface at fixed $r$ and time $t$ is $4\pi f^2(r)$, then $f(r)$ is the areal radius. The feature of the Euclidean metric is that $f(r)$ coincides with the proper radius,  $\int^r dr' f'(r')$.  

For all these models Condition II is satisfied, i.e. $R^\star_{\mu\nu}$ is Einstein. Since eq.\eqref{RstarStar} gives $R^\star =0$, it is $R^\star_{\mu\nu}=0$.
%The electric tensor is:
%$$E(r)=\frac{n-3}{n-2}\frac{1}{f'^2}\left[  \frac{b'}{b}\frac{d}{dr} \log (f'f) 
% - \frac{b''}{b}  \right ]  $$
\end{example}

\section{Friedmann equations in spherical GRW spacetimes}
Spherical GRW spacetimes are characterized by $\dot u=0$ or, equivalently, by the metric 
\begin{align}
 ds^2 = -dt^2 + a^2(t) [f_1(r)^2 dr^2 + f_2^2(r) d\Omega_{n-2}^2] \label{spherGRW}
 \end{align}
With $b=1$, the dot coincides with the time derivative.
The Einstein equations with anisotropic fluid source \eqref{Tij} give $q_i$=0 (no heat current) and
\begin{align*}
\mu=-\xi +\tfrac{1}{2}R ,  \quad p_r+(n-2)p_\perp= -\xi  - \tfrac{1}{2}(n-3)R , \quad p_r-p_\perp = - (n-2) \frac{E(r)}{a^2(t)}  
\end{align*}
It is the electric component $E(r)$ of the Weyl tensor, eq.\eqref{E(r)}, that breaks isotropy. With \eqref{Rcov}, $\varphi =\dot a /a$ (the Hubble parameter), 
and $\xi =(n-1)\ddot a/a$, the equations for the energy density and the radial pressure are
\begin{align}
&\mu =  \frac{R^\star (r)}{2a^2}  + \frac{1}{2}(n-1)(n-2)\frac{\dot a^2}{a^2}\\
&(n-1)p_r+(n-3)\mu =   - (n-1)(n-2)\frac{\ddot a}{a} - (n-2)^2 \frac{E(r)}{a^2}
\end{align}
The equations for $\mu$ is equal to the first Friedmann equation in RW cosmology. It shows that $a^2 \mu$ is the sum
of a function of $r$ and a function of time. The derivative of $2a^2 \mu $ and the equation for
$p_r$ give:
\begin{align}
\dot \mu =& - \frac{\dot a}{a} \left [ (n-1)(p_r+\mu) + (n-2)^2 \frac{E}{a^2}\right ]= - \frac{\dot a}{a} (n-1)(P+\mu) 
\end{align}
where $P$ is the effective pressure.\\
In a spherical GRW spacetime, the expressions for $R^\star $ and $E(r)$ give
$$ \frac{R^\star(r)}{n-2} + 2\frac{n-2}{n-3} E(r) = \frac{n-1}{f_2^2}\left [1- \frac{f_2^{\prime 2}}{f_1^2}\right ] $$
It is simple to check that in a spherical GRW spacetime, the Euclidean condition $f_1=f_2'$ (example \ref{EUCL}) 
is necessary and sufficient for $R^\star=0$ and $E=0$ (the spacetime is RW with flat space).
\begin{example}
In ref.\cite{Cadoni20} the authors discuss a cosmological model with a non-radiative anisotropic fluid for an effective description of barionic matter and dark
energy without assuming the presence of dark matter. The pressure anisotropy is the source of the small-scale inhomogeneities of the universe. The metric
in cosmic time is \eqref{spherGRW} with $f_1^2(r)=1/f(r)$ and $f_2^2(r)=r^2$. It is found:
$$ R^\star = 2\frac{1-f}{r^2} - (n-2)\frac{f'}{r} , \quad E(r)=\frac{n-3}{n-2}\left [ \frac{1-f}{r^2} +\frac{f'}{2r}\right ]$$ 
In particular, the following function is used $ f(r) =1 - \frac{2G}{r}[m_B(r) + m]$
where $m_B$ is the Misner-Sharp mass for the inhomogeneous distribution of baryonic matter. \\
The choice $E(r)=0$ implies $f(r)=1-k r^2$, and the spacetime is RW with curvature $k$.
\end{example}

\section{Spherical DW models in $f(R)$ gravity} 
A generalisation of Einstein's theory are the $f(R)$ theories. In such theories, the scalar $R$ in the gravitational action is replaced by a smooth function $f(R)$. They were introduced by Buchdahl in 1970 \cite{Buchdahl70} and gained popularity with the works by Starobinsky on cosmic inflation \cite{Starobinsky80}. Now
they are explored as possible cosmological theories for a geometric description of effects that otherwise require
the introduction of dark matter and dark energy  in the energy-momentum tensor. \\
Variation with respect to $g_{ij}$, modulo surface terms, gives the field equations:
\begin{align} 
& f'(R)R_{kl}  - [f^{'''}(R) (\nabla_k R)(\nabla_l R) + f^{''}(R) \nabla_k\nabla_l R] \label{fieldeqs2}\\
& + g_{kl} [ f^{'''}(R) (\nabla_k R)^2 + f^{''}(R) \nabla^2 R-\tfrac{1}{2} f(R)]=
\kappa T_{kl} \nonumber
\end{align}
%where a prime denotes derivative with respect to the argument. 
In a GRW spacetime of dimension $n$, if the Weyl tensor is harmonic, then $R_{jk}$ is rank-1 (perfect-fluid) and all curvature corrections in \eqref{fieldeqs2} 
are of the same form. Therefore, they can be intrepretated as geometric corrections to the perfect fluid parameters of the fluid matter-radiation source \cite{Capozziello19}. A systematic study is done in \cite{Capozziello2022} with the notion of {\em perfect scalars} to prove the status of 
perfect fluid of corrections in various extended gravity theories.

$f(R)$ gravity is also being considered to model stellar collapse  \cite{Goswami14}\cite{Capozziello11}\cite{Chakrabarti18}\cite{Chakrabarti16}\cite{Jaryal21}. 
Since the exterior solution is pure radiation ($T^k{}_k=0$), the $f(R)$ equation is still solved by the Vaidya metric, where $R=0$. 
The junction conditions with the inner solution have the
additional contraints $R=0$ and $N^j\partial_j R=0$ at the boundary, that exclude some inner metrics \cite{Senovilla13}.

We show that in spherical DW spacetimes the geometric terms in the field equations \eqref{fieldeqs2} have the same tensor form as in \eqref{Tij}. 
Therefore, for an anisotropic fluid with energy-momentum tensor \eqref{Tij}, the effect of $f(R)$ gravity is only to 
modify the parameters $\mu$, $p_r$, $p_\perp$ and $q_j$ of the fluid. If the input fluid tensor is isotropic, the geometric $f(R)$ 
corrections may produce anisotropy.

We introduce the following: 
\begin{definition}
On a spherical DW spacetime, a scalar $X$ is a {\em perfect 2-scalar} if
\begin{align} 
\nabla_j  X =  -u_j \dot X + \dot u_j  \tilde X
\end{align}
It follows that $\dot X= u^k \nabla_k X$ and $\tilde X= \frac{1}{\eta}\dot u^k\nabla_k X$.\\
A linear combination of the tensors $u_iu_j$, $g_{ij}$, $u_i\dot u_j$ and $\dot u_i \dot u_j$ is a rank-2 perfect tensor (also named
2-Einstein tensor \cite{Shaikh22}).
\end{definition}
\noindent
Perfect 2-scalars are $\varphi $ (see eq.\eqref{U2}), $\eta $ (see eq.\eqref{nablaeta} with $w_k=0$), $\xi $ (see eq.\eqref{nablaxi}). 
The Ricci and the electric tensors are rank-2 perfect tensors.\\
If $X$, $Y$ are perfect 2-scalars, also $X+Y$ and
$XY$ are such (Leibnitz rule). 
\begin{remark}
An equivalent characterization is $N_j{}^k \nabla_k X=0 $,
where $N_{jk}$ is the projection \eqref{projector}. 
Since in spherical components $N_{tt}=N_{rr}=N_{rt}=0$, a scalar function is a perfect 2-scalar if
and only if it depends only on $t$ and $r$.
\end{remark}
\begin{lem} If $X$ is a perfect 2-scalar, then $\dot X$ and $\tilde X$  are perfect 2-scalars, and $\dot{\tilde X} = \dot X +\tilde{\dot X}$.
\begin{align}
\nabla_i \dot X  = -u_i \ddot X + \dot u_i (\dot{\tilde X}-\dot X) \label{STAT1}\\
 \nabla_i\tilde X= - u_i (\dot X+\tilde{\dot X} )+  \dot u_i \tilde{ \tilde X} \label{STAT2}
\end{align}
\begin{proof} Consider the Hessian $\nabla_i\nabla_j X =-\dot X \nabla_i u_j + \tilde X \nabla_i\dot u_j - u_j \nabla_i \dot X
+\dot u_j \nabla_i \tilde X$. Exchange the indices and subtract, use the fact that $\dot u_i$ is closed:
$$0=-\dot X(-u_i \dot u_j +u_j \dot u_i) -u_j \nabla_i \dot X + u_i\nabla_j \dot X + \dot u_j \nabla_i\tilde X -\dot u_i\nabla_j \tilde X$$
Contraction with $u^j$ gives the first statement.
Contraction with $\dot u^j$ gives: $0=\dot X \eta u_i   + u_i\dot u^j\nabla_j \dot X + \eta \nabla_i\tilde X -\dot u_i \dot u^j\nabla_j \tilde X$. Then the second statement follows. Contraction of $u^i$ with \eqref{STAT2} gives the last relation.
\end{proof}
\end{lem}
\begin{prop}
The Hessian of a perfect-2 scalar is a rank-2 perfect tensor:
\begin{align}
\nabla_i\nabla_j X =& -\dot X \varphi g_{ij}   
+u_iu_j (\ddot X -\varphi\dot X -\eta\tilde X)- (\dot u_i u_j +u_i \dot u_j)(\tilde{\dot X}-\varphi\tilde X)\\ 
&+  \dot u_i \dot u_j (\tilde{ \tilde X} + \tilde X \frac{\tilde\eta}{2\eta^2}) +   N_{ij}  \tilde X\frac{-\eta + \nabla_r \dot u^r - \frac{1}{2\eta} \tilde\eta}{n-2} \nonumber
\end{align}
\begin{proof}
$\nabla_i\nabla_j X = - \dot X \nabla_i u_j + \tilde X \nabla_i\dot u_j - u_j \nabla_i \dot X
+\dot u_j \nabla_i \tilde X = -\dot X (\varphi u_i u_j + \varphi g_{ij} -u_i\dot u_j)  + \tilde X \nabla_i\dot u_j 
-u_j [ -u_i \ddot X + \dot u_i (\dot{\tilde X}-\dot X)] +\dot u_j [- u_i (\dot X+\tilde{\dot X} )+  \dot u_i \tilde{ \tilde X}] =
 -\dot X \varphi g_{ij}  + \tilde X \nabla_i\dot u_j 
+u_iu_j (\ddot X -\varphi\dot X)- (\dot u_i u_j +u_i \dot u_j)\tilde{\dot X} +  \dot u_i \dot u_j\tilde{ \tilde X}$.\\
 For a spherical DW spacetime $\nabla_i \dot u_j$ is given by eq.\eqref{nabladot} with $\Pi_{ij}=0$ and $w_j=0$, and
%\begin{align}
%\nabla_j \dot u_k = -\eta u_j u_k +\varphi (u_j \dot u_k + \dot u_j u_k) + \dot u_j \dot u_k \frac{\tilde\eta}{2\eta}+ N_{jk} \frac{N_{pq} \nabla^p \dot u^q}{n-2} 
%\label{NABDOTU}
%\end{align}
it is a rank-2 perfect tensor.
%$N_{rs} \nabla^r \dot u^s = [u_r u_s+ g_{rs} - \frac{\dot u_r \dot u_s}{\eta} ] \nabla^r \dot u^s = - u_r \dot u_s \nabla^r u^s +\nabla_r \dot u^r - \frac{1}{2\eta} \tilde\eta = -\eta + \nabla_r \dot u^r - \frac{1}{2\eta} \tilde\eta$.
\end{proof}
\end{prop}
The curvature scalar $R$ of a spherical DW spacetime is a function only of $r$ and $t$, then it is a perfect 2-scalar
and the Hessian $\nabla_i\nabla_j R$ as well as the product $\nabla_i R \nabla_j R$ are rank-2 perfect tensors.
We then conclude:
\begin{prop}
On a spherical DW spacetime, the $f(R)$ field equations imply an energy-momentum tensor \eqref{Tij} that is a rank-2 perfect tensor.
\end{prop}

\begin{example} In Wagh's metric \eqref{Waghmetric} 
%the Ricci tensor \eqref{RicciSphere} is isotropic,
%\begin{align*}
%R_{jk} = u_j u_k (-\xi+\nabla_p\dot u^p)
%+ \frac{h_{jk}}{n-1} (R-\xi  + \nabla_p \dot u^p) - (n-2)\varphi (u_j \dot u_k + \dot u_j u_k)
%\end{align*} 
%with $R$, $\xi$ given in example~\ref{EXWAGH}. 
the following terms in  \eqref{fieldeqs2} are evaluated:
\begin{align}
\nabla_k R=& -\dot R u_k -2R \dot u_k \label{nablaRWAGH}\\
\nabla_l \nabla_k R =& (\ddot R-\varphi \dot R) u_k u_l - (\varphi \dot R +2R\eta) g_{kl} \\
&+ (3\dot R -2R\varphi) (u_k \dot u_l + \dot u_k u_l) + 8R \dot u_k \dot u_l   \nonumber
% \nabla^k\nabla_k R =& -\ddot R-(n-1)\varphi \dot R -2(n-4)R\eta.
\end{align}
\begin{proof} Being $R$ a perfect 2-scalar: $ \nabla_k R = - \dot R u_k + (\frac{1}{\eta} \dot u^r \partial_r R)\dot u_k $. 
The curvature scalar $R$ is \eqref{RWagh}, then $\partial_r R = -2(f'_2/f_2)R$. In spherical coordinates:
$$ \frac{1}{\eta} \dot u^r \partial_r R = \frac{b^2 a^2 f_1^2}{b_r^2}\frac{b^r}{b} \left (-2\frac{f'_2}{f_2}R\right ) = -2 
 \frac{b}{b_r}\frac{f'_2}{f_2}R =-2R $$
$\nabla_l \nabla_k R = -(\nabla_l \dot R)u_k - \varphi \dot R h_{kl} + 3 \dot R u_l \dot u_k +4 R \dot u_l \dot u_k -2R \nabla_l \dot u_k$. Since $\dot u_k$
is closed in DW spacetimes, the antisymmetric part is: $0=-u_k \nabla_l \dot R + u_l \nabla_k \dot R + 3 \dot R(u_l\dot u_k - u_k \dot u_l)$. 
Contraction with $u^k$ gives
$\nabla_l \dot R = -\ddot Ru_l - 3\dot R \dot u_l$. This is inserted in the Hessian:\\
$ \nabla_l \nabla_k R = (\ddot R-\varphi \dot R) u_k u_l - \varphi \dot R g_{kl} + 3\dot R (u_k \dot u_l + \dot u_k u_l) + 4 R \dot u_k \dot u_l 
- 2R\nabla_l \dot u_k$
The last term is \eqref{nabladot}. In Wagh's metric it is $\dot u^k\nabla_k \eta = -2\eta^2$ and $\nabla_p \dot u^p = (n-2)\eta$. 
Eq. \eqref{Nnabladot} is $N_{pq}\nabla^p \dot u^q= \nabla_p \dot u^p$ and the result follows.
\end{proof}
While the Ricci tensor $R_{ij}$ is isotropic, the $f(R)-R$ terms in \eqref{fieldeqs2} add a contribution
$\dot u_i \dot u_j [-8Rf''(R) - 4R^2 f'''(R)]$
that implies fluid anisotropy. This term is zero only for $f(R) = R + K \log R$. However $K\neq 0$ violates the condition $f'(0)$ to be a finite constant,
that guarantees a finite radiation energy density in the Vaidya solution \cite{Senovilla13}. In conclusion, full isotropy in Wagh's metric is only
possible in Einstein gravity, $f(R)=R$.

For a collapsing star, the boundary condition $R=0$ at $r_\Sigma $ also satisfies the boundary condition $\dot u^k\nabla_k R =0$ because of \eqref{nablaRWAGH}.\\
Wagh's metric has been investigated with $f(R)=R+\alpha R^2$ in \cite{Chowdhury20} and $f(R)=\xi R^4$ in \cite{Solanki22}. To satisfy the boundary
condition \eqref{BEQ}, they set to zero the prefactor, i.e. $1/f_2(r_\Sigma)=0$. This also makes $R=0$ at the boundary, but implies divergence
of the metric tensor, making the relation \eqref{CONTIN} among proper time $\tau$, the Vaidya parameter $v$ and the stellar time $t$ unclear. 
On the other hand, 
a finite value $f_2(r_\Sigma)$ makes the conditions \eqref{BEQ} and $R=0$ two incompatible equations for $a(t)$. Wagh's metric is too constrained for $f(R)$ stellar collapse.
\end{example}

\subsection{Spherical GRW spacetimes in $f(R)$ gravity} The missing acceleration $\dot u_i$ is replaced by the radial unit vector $N_i$, with
$N_0=0$ and $N_\mu=a f_1 \delta_{r\mu}$ in coordinates $(t,r,\theta_a)$. The covariant expression holds:
\begin{align} \nabla_i N_j = \frac{f_2' }{af_1f_2} (g_{ij} +u_i u_j - N_i N_j) + \varphi N_i u_j 
\end{align}
where $\varphi =\dot a/a$. Any function of $t$ and $r$ has gradient $\nabla_j F= -u_j  \dot F+ \frac{N_j}{a f_1}F'$, 
where $\dot F$ and $F'$ are partial derivatives in $t$ and $r$. Then:
\begin{align}
\nabla_i \nabla_j F =& -(\nabla_i u_j)\dot F -u_j \nabla_i \dot F  + \nabla_i (\frac{N_j}{af_1})F' + \frac{N_j}{af_1}  \nabla_i F'\\
=& u_i u_j  \ddot F + h_{ij} \left[ \frac{f_2'}{a^2 f_1^2 f_2} F'   -\varphi \dot F \right ] + \frac{N_iN_j }{a^2 f_1^2}
\left [ F'' - \left( \frac{f_1'}{f_1}+ \frac{f_2'}{f_2}\right )F' \right ] \nonumber\\
& + \frac{u_i N_j + u_j N_i}{af_1} ( \varphi F' - \dot F' ) \nonumber
\end{align}
The Hessian has the form of a perfect rank-2 tensor with basic vectors $u_i$ and $N_i$. The curvature scalar \eqref{Rcov}
depends on $r$ through $R^\star $ and on $t$ through $a$:
$$ R' = \frac{R^{\star\prime}}{a^2}, \qquad \dot R = - 2\frac{R^\star}{a^2}\varphi  +2 (n-1)(n-2)\varphi \dot\varphi +2\dot \xi $$
The Hessian $\nabla_i\nabla_j R$ and $\nabla_i R\nabla_j R$ are rank-2 tensors built with $u_i$ and $N_i$.  
Therefore, in spherical GRW spacetimes, $f(R)$ produces heat-like and anisotropic geometric terms in the field equations.
Such terms are absent if $R^\star $ is a constant.

%In \cite{Chakrabarti16} the homothetic case $\varphi^2+\dot\varphi=0$ is considered, i.e $a_{tt}=0$. Then $a(t) = C(t-t_0)$.
%\end{example}
\section{Conclusions}
Spherical doubly warped spacetimes are tractable settings for models of stellar collapse, inhomogeneous cosmology, and wormholes. 
In general, a DW spacetime is characterised by the existence of a shear, vorticity-free velocity $u_j$ and its orthogonal
acceleration field $\dot u_j$ (the latter is closed, and the expansion parameter only changes in the two directions). 
If $\dot u_j=0$ the spacetime is Generalized Robertson-Walker. The presence of two vectors makes the geometry rich. 

We obtained properties for the Weyl tensor and the covariant structure of the Ricci tensor on general DW spacetimes, that extend known 
results for GRW, and an identity for the Ricci and electric tensors in the space sub-manifold. \\
In spherical symmetry, the space sub-manifold is conformally flat, $R_{ij}$, $\nabla_i \dot u_j$ and $E_{ij}$ are rank-2 tensors
built with $u_i$ and $\dot u_i$.\\
The genaral structure of the Ricci tensor dictates that of the energy-momentum 
tensor of an anisotropic fluid in the Einstein equations. The deviation from the perfect fluid form is
determined by the electric tensor $E_{ij}$ and the gradient of the acceleration $\nabla_i \dot u_j$. In the Einstein equations they yield 
the general pressure anisotropy term, the general heat current with $q_j$ proportional to the acceleration,
in a spherical DW metric. The ensuing explicit expressions for the energy density, heat flux, 
radial and transverse pressures are displayed. \\
We then focused on conditions that make these models isotropic, reproduced the junction conditions by 
Santos for star collapse, obtained the Misner-Sharp mass, 
and the equality of radial pressure with the modulus of the heat current. 
Several features investigated in literature on stellar collapse are here framed in a simple manner.\\ 
The form of the Ricci tensor can be exploited to study expansion-free compact objects, or the
anisotropy in dissipative cosmology. We obtained the Friedmann equations in spherical GRW space-times, 
with a departure from standard FRW cosmology.\\
Finally, we introduced the useful concept of perfect 2-scalars, to show that the $f(R)$ corrections to Einstein gravity
amount to terms with the same tensor form as the energy-momentum of the fluid source.

\subsection*{Data availability}
Data sharing not applicable to this article as no datasets were generated or analysed during the current study.

\section{APPENDIXES}
\subsection{Proof of Prop.~\ref{DOTRetc}}\label{PROOFDOTR}
%\begin{proof}
$\nabla_k\xi =(n-1)\nabla_k(\varphi^2 +u^l \nabla_l \varphi) = (n-1)[ 2\varphi \nabla_k\varphi + (\nabla_k u^l)\nabla_l \varphi + u^l\nabla_l\nabla_k
\varphi ]$. Now use \eqref{U1} and \eqref{U2}:  $\nabla_k\xi = (n-1)[ 2\varphi ( -u_k \dot \varphi  - \varphi \dot u_k) + (\varphi u_k u^l +\varphi \delta_k^l
- u_k\dot u^l)(-u_l \dot \varphi  - \varphi \dot u_l) + u^l\nabla_l ( -u_k \dot \varphi  - \varphi \dot u_k)] $. After simplifications:
$\nabla_k\xi = -(n-1)( 2u_k\varphi  \dot \varphi  +3\varphi^2  \dot u_k +2\dot u_k \dot \varphi  + u_k \ddot \varphi -\varphi\eta u_k+\varphi \ddot u_k)$. 
Contraction of \eqref{nabladot} with $u^j$ gives $\ddot u_k = \eta u_k - \varphi \dot u_k$. Then:
$\nabla_k\xi = -(n-1)( 2\varphi  \dot \varphi  + \ddot \varphi )u_k -2(n-1)(\varphi^2  +\dot \varphi )\dot u_k $. This is eq.\eqref{nablaxi}.

$R_{km}\dot u^m = g^{jl} (\nabla_j \nabla_k \dot u_l -\nabla_k \nabla_j \dot u_l ) =
\nabla_j \nabla_k \dot u^j - \nabla_k \nabla_j\dot u^j $. The contraction with $u^k$ gives:
$u^k\nabla_k \nabla_j\dot u^j = -R_{km}u^k \dot u^m +u^k \nabla_j \nabla_k \dot u^j = -(n-2)\varphi \eta - (\nabla_j u^k)(\nabla_k \dot u^j) + \nabla_j \ddot u^j
=  -(n-2)\varphi \eta -\varphi u_j \ddot u^j - \varphi \nabla_p\dot u^p + u_j \dot u^k \nabla_k \dot u^j + \nabla_j (\eta u_j -\varphi \dot u_j)$. Now use the fact that $\dot u$ is closed: 
$u^k\nabla_k \nabla_j\dot u^j =  -(n-2)\varphi \eta -\varphi u_j \ddot u^j - \varphi \nabla_p\dot u^p + \frac{1}{2} \dot \eta + \dot \eta + (n-1)\varphi\eta - \dot u^j \nabla_j\varphi - \varphi \nabla_p \dot u^p
= \varphi \eta -\varphi u_j (\eta u^j -\varphi \dot u^j) - 2\varphi \nabla_p\dot u^p + \frac{3}{2} \dot \eta   - \dot u^j (-\dot\varphi u_j -\varphi \dot u_j)
= 3\varphi \eta  - 2\varphi \nabla_p\dot u^p + \frac{3}{2} \dot \eta $. Finally use \eqref{nablaeta} to obtain $\dot \eta = -2\varphi \eta$. Then: $u^k\nabla_k \nabla_j\dot u^j =  - 2\varphi \nabla_p\dot u^p$. This is \eqref{dotnabladotu}.

The divergence of \eqref{Ricciu} is $\frac{1}{2}\dot R + R_{km} (\nabla^k u^m)=u^k\nabla_k( \xi - \nabla_p\dot u^p) + (\xi -\nabla_p\dot u^p)\nabla^k u_k + (n-2)\nabla_k (\varphi \dot u^k)$. Simplify and obtain:
$ \dot R -2\dot \xi =-2\varphi (R+2\nabla_p\dot u^p -n\xi) - 2u^k\nabla_k(\nabla_p\dot u^p)$. The last term has just been evaluated and simplifies 
the formula to give \eqref{dotR}.

\subsection{Proof of equation \eqref{RicciqE}}\label{INTEGRABILITY}
Let $h^\star_{\mu\nu}=g^\star_{\mu\nu} -n_\mu n_\nu$. We use  $\partial_\nu\Theta = n_\nu (n^\rho\partial_\rho \Theta)$
(Prop. \ref{nablanf1f2}).
\begin{align*}
(n-2) R^\star_{\mu\nu\rho\sigma}n^\sigma =&(n-2) (\nabla^\star_\mu \nabla^\star_\nu - \nabla^\star_\nu \nabla^\star_\mu )n_\rho\\
=& \nabla^\star_\mu  \left[\Theta (g^\star_{\nu\rho} - n_\nu n_\rho) \right ] - \nabla^\star_\nu  \left[\Theta (g^\star_{\mu\rho} - n_\mu n_\rho) \right ]\\
 =& (h^\star_{\nu\rho} \partial_\mu \Theta-h^\star_{\mu\rho}\partial_\nu\Theta) - \Theta (n_\nu \nabla^\star_\mu n_\rho
 - n_\mu \nabla^\star_\nu n_\rho ) \\
= & (h^\star_{\nu\rho} \partial_\mu \Theta-h^\star_{\mu\rho}\partial_\nu\Theta) - \Theta^2 (n_\nu g^\star_{\mu\rho}
 - n_\mu g^\star_{\nu\rho} )/(n-2) \\
 =&\xi^\star (g^\star_{\mu\rho} n_\nu -g^\star_{\nu\rho}n_\mu)
 \end{align*}
with $\xi^\star $ given by \eqref{xistar}. The contraction with the metric tensor $g^{\star\mu\rho}$ shows that it is an eigenvalue: $R^\star_{\nu\sigma}n^\sigma =\xi^\star n_\nu $.
 Since $C^\star_{\mu\nu\rho\sigma}=0$ we have:
 \begin{align*}
 0=R^\star_{\mu\nu\rho\sigma}+ \frac{g^\star_{\mu\sigma} R^\star_{\nu\rho}-g^\star_{\nu\sigma}R^\star_{\mu \rho} +R^\star_{\mu \sigma}g^\star_{\nu \rho}-R^\star_{\nu\sigma} g^\star_{\mu \rho}}{n-3} - R^\star\frac{g^\star_{\mu \sigma}g^\star_{\nu \rho}-g^\star_{\nu \sigma}g^\star_{\mu \rho}}{(n-2)(n-3)}
 \end{align*}
 The contraction with $n^\mu n^\sigma$
 \begin{align*}
 0=-\xi^\star (g^\star_{\nu\rho}- n^\mu n^\rho)+ \frac{n-2}{n-3}[ R^\star_{\nu\rho}- n_\nu n_\rho \xi^\star +\xi^\star g^\star_{\nu \rho}-\xi^\star n_\nu n_\rho ] - R^\star\frac{g^\star_{\nu \rho}-n_\nu n_\rho}{n-3}
 \end{align*}
and the Ricci tensor $R^\star_{\nu\rho}$ in \eqref{RicciqE} is obtained.
\subsection{Proof of Lemma \ref{LEMMAACC}}\label{APPLEMMAACC}
We need: $ \nabla_p\dot u^p = \Gamma_{0\mu}^0 \frac{b^\mu}{b} + \nabla_\mu (\frac{b^\mu}{b}) = 
\frac{b_\mu b^\mu}{b^2}+ \frac{1}{b} \nabla_\mu b^\mu -
\frac{b_\mu b^\mu}{b^2} = \frac{1}{a^2(t)b}\nabla_\mu^\star b^\mu $. Then:
\begin{align}
\nabla_p\dot u^p =  \frac{1}{a^2(t)b}\nabla_\mu^\star b^\mu. \label{nablapdotup}
\end{align}
In the frame \eqref{F1F2}: 
\begin{align*}
0 &=  \nabla_p \dot u^p - (n-1)\left( \frac{\dot u^i\nabla_i \eta}{2\eta}+\eta\right) \\
& = \nabla_p\dot u^p - (n-1) \left( g^{rr} \frac{\partial_r b}{b} \partial_r \log \sqrt \eta +\eta \right )\\
& = \frac{1}{a^2(t)b} \nabla^\star_\mu b^\mu - (n-1) \left(  \frac{1}{a^2 f_1^2}\frac{\partial_r b}{b} \partial_r \log \left(\frac{\partial_r b}{f_1 b}
\right) + \frac{1}{a^2(t)} \frac{(\partial_r b)^2}{f_1^2 b^2}  \right )\\
&= \frac{1}{a^2(t) f_1^2 b}\,\left [ f_1^2 \nabla^\star_\mu b^\mu -(n-1)\left( \partial_r b \, \partial_r \log 
\left(\frac{\partial_r b}{f_1 b} \right) + \frac{(\partial_r b)^2}{b}  \right ) \right ] \\
&= \frac{1}{a^2(t) f_1^2 b}\,\left [ f_1^2 \nabla^\star_\mu b^\mu  -(n-1)(\partial^2_r b - \partial_r b\, \partial_r \log f_1) \right ]
\end{align*}
Now we use eq.\eqref{divdotu}: $f_1^2  \nabla^\star_\mu b^\mu  = \partial^2_r b -   (\partial_r b)\partial_r \log (f_1 f_2)
 + (n-1)  (\partial_r b) (\partial_r \log f_2)$ and obtain the result.
%
%\subsection{Proof of Condition II}\label{COND2} 
%Eq.\eqref{44} is a Bernoulli equation for $f_1(r)$: 
%\begin{align*}
%f'_1 + A(r) f_1 +B(r)f_1^3 =0 \quad\text{with}\;
% A(r) = - \frac{d}{dr} (\log \frac{d}{dr}\log f_2), \; B(r) =  -\frac{1}{f_2 f_2'}
%\end{align*}
%Divide by $f_1^3$ and set $y=1/f_1^2$. The equation now is: $ y' - 2y A - 2B =0$. 
%According to the general solving formula:
%\begin{align*}
% \frac{1}{f_1(r)^2} =
 %= e^{2\int^r dr' A(r')} \left [ \int^r dr' 2B(r') e^{-2 \int^r dr' A(r')} +C \right ]=
% -[\frac{d}{dr}\log f_2(r) ]^{-2}\left [ \int^r dr' \frac{2}{f_2 f_2'} \frac{f_2'(r')^2}{f_2(r')^2} +C \right ]
%= -\frac{f_2(r)^2}{f'_2(r)^2} \left [ \int^{f_2}  \frac{2 df }{f^3} +C \right ] = 
%\frac{1+C f_2(r)^2}{f'_2(r)^2}  \end{align*}

\subsection{A new class of solutions of eq.\eqref{EQBANERJEE}}\label{SOLLM}
With $B=b_x/b$ and $G=F_x/F = - f_{1x}/f_1$, it is $b_{xx}/b= B_x+B^2$ and $F_{xx}/F=G_x+G^2$. Eq.\eqref{EQBANERJEE} becomes:
$B_x -(n-3) G_x= -B^2 -2BG + (n-3)G^2 $. 
We search solutions such that $B=(\alpha -1) G -\beta$. Then:
$(n-2 -\alpha ) G_x = [\alpha^2 - (n-2)]G^2 - 2\beta \alpha G +\beta^2 $. A first integral is
\begin{gather*}
\frac{d}{dx} \log \Big | \frac{G-g_+}{G-g_-}\Big | = \frac{2\beta\sqrt{n-2}}{(n-2)-\alpha }\quad \text{where}\; 
g_\pm = \beta \frac{\alpha \pm \sqrt{n-2}}{\alpha^2 - (n-2)}\\
 G(x) =  \frac{g_+  \pm g_-  K e^{\nu x}}{1\pm K e^{\nu x}} \qquad \nu = \frac{2\beta \sqrt{n-2}}{(n-2)-\alpha } , \; K\ge 0
%-\frac{d}{dx}\log f_1=\beta \frac{\alpha-\sqrt{n-2}}{\alpha^2 - (n-2)} - \frac{2\beta}{\nu} \frac{\sqrt{n-2}}{\alpha^2 - (n-2)} 
%\frac{d}{dx} \log[e^{-\nu x} \pm K]
\end{gather*}
Integrate $G=-f_{1x}/f_1$ and put $x=r^2$ in the end. Once $f_1$ is obtained, $b=C'e^{-\beta r^2} f_1^{1-\alpha}$.
The result are the pairs:
\begin{align}
f_1(r)=& e^{-\beta r^2 \frac{1}{\alpha - \sqrt{n-2}}} (1 \pm Ke^{\nu r^2})^{-\frac{\alpha -(n-2)}{\alpha^2 - (n-2)}}\\
b(r) =& e^{-\beta r^2 \frac{1-\sqrt{n-2}}{\alpha - \sqrt{n-2}}} (1 \pm Ke^{\nu r^2})^{-(1-\alpha)\frac{\alpha -(n-2)}{\alpha^2 - (n-2)}}
\end{align}

\subsection{The doubly warped metric}\label{DWcoeffs}
%Christoffel symbols and Ricci tensor for the {\em doubly warped} metric 
$ ds^2 = -b^2({\bf x}) dt^2 + a^2(t) g^\star_{\mu\nu} ({\bf x}) dx^\mu dx^\nu $.\\
Notation: $a_t=\partial_t a$, $b_\mu =\partial_\mu b$, $b^\mu = g^{\star\mu\nu}b_\nu$.\\
\noindent
$\bullet$ {\sf Metric tensor:} 
\begin{align} 
g_{00}=-b^2({\bf x}), \quad g_{\mu\nu}=a^2(t)g^\star_{\mu\nu}({\bf x}),\quad g^{00} =-\frac{1}{b^2},\quad g^{\mu\nu}=
\frac{1}{a^2}g^{\star\mu\nu}
\end{align}
$\bullet$ {\sf Christoffel symbols:} $\Gamma_{ij}^k = \tfrac{1}{2}g^{k\ell}(\partial_i g_{j\ell} +\partial_j g_{i\ell} - \partial_\ell g_{ij}) $.
\begin{align}
&\Gamma_{00}^0=0, \quad \Gamma_{\mu 0}^0 =\frac{b_\mu}{b},\quad \Gamma_{00}^\mu =\frac{bb^\mu}{a^2},\quad
\Gamma_{\mu 0}^\lambda =\delta^\lambda_\mu \frac{a_t}{a},\\
&\Gamma_{\mu\nu}^0 = \frac{aa_t}{b^2} g^\star_{\mu\nu}, \quad \Gamma_{\mu\nu}^\lambda = \Gamma_{\mu\nu}^{\star\lambda}
\end{align}
$\bullet$ {\sf Riemann tensor:} $R_{jkl}{}^m = -\partial_j \Gamma^m_{k,l} + \partial_k \Gamma^m_{j,l} + \Gamma_{j,l}^p\Gamma^m_{kp} - \Gamma_{k,l}^p \Gamma_{jp}^m $
\begin{align}
R_{\mu 0\rho}{}^0 &= -\frac{1}{b} \nabla^\star_\mu b_\rho + \frac{a a_{tt}}{b^2} g^\star_{\mu\rho},  \qquad
R_{\mu\nu\rho}{}^0  = \frac{a a_t}{b^3} (b_\mu g^\star_{\nu\rho} -  b_\nu g^\star_{\mu\rho}) \\
R_{\mu\nu\rho}{}^\sigma &= \, R^\star_{\mu\nu\rho}{}^\sigma + \frac{a_t^2}{b^2} (g^\star_{\mu\rho}\delta^\sigma_\nu - g^\star_{\nu\rho}\delta^\sigma_\mu) 
\end{align}
$\bullet$ {\sf Ricci tensor:} $R_{ij} =\partial_m \Gamma_{ij}^m -\partial_i \Gamma^m_{m j}
+\Gamma_{ij}^k \Gamma_{km}^m -\Gamma_{im}^k \Gamma_{kj}^m$  %2.41
\begin{align}
&R_{00}= \frac{b}{a^2}\nabla^\star_\mu b^\mu -(n-1)\frac{a_{tt}}{a} \label{R00}\\
&R_{\mu 0}= (n-2) \frac{a_t}{a} \frac{b_\mu}{b} \label{Rmu0}\\
&R_{\mu\nu} = R^\star_{\mu\nu} + g^\star_{\mu\nu}  \frac{1}{b^2} [ (n-2)a_t^2  + aa_{tt} ]-\frac{1}{b} \nabla^\star_\mu  b_\nu
\label{RicciSymb}
\end{align}
$\bullet$ {\sf Scalar curvature:} $R=g^{ij}R_{ij}=-\frac{1}{b^2}R_{00} +\frac{1}{a^2}\sum_\mu R_{\mu\mu}$
\begin{align}
R=\frac{R^\star}{a^2} -\frac{2}{a^2b}\nabla^\star_\mu b^\mu + (n-1)\frac{1}{a^2b^2}[(n-2)a_t^2 +2a a_{tt}] \label{RRstar}
\end{align}

\subsection{Radial space metric, d=1+N}\label{SPHEREMETRIC} $ dx^2 =  f_1^2(r)dr^2 + f_2^2(r) \sum_{a=1}^N g_a^2(\boldsymbol \theta ) d\theta_a^2 $
\begin{gather*}
 %dx^2 =  f_1^2(r)dr^2 + f_2^2(r) \sum_{a=1}^N g_a^2(\boldsymbol \theta ) d\theta_a^2 \\
\Gamma^{\star r}_{rr} = \partial_r \log f_1, \quad \Gamma^{\star r}_{r\theta}=0, \quad \Gamma^{\star a}_{rr}=0,
\quad  \Gamma^{\star r}_{a a'}=
- \delta_{aa'}g^2_a(\boldsymbol \theta) \frac{f_2^2}{f_1^2}  (\partial_r \log f_2)\\
\Gamma^{\star a'}_{r a} =\delta_{aa'} \partial_r \log f_2, \quad 
\Gamma^{\star a}_{a' a''} =\hat\Gamma^{a}_{a' a''}\; \text{are independent of $r$}
%=(\delta_{\theta\theta''} \partial_{\theta'}  + \delta_{\theta\theta'} \partial_{\theta''}) \log d_\theta -
%\delta_{\theta'\theta''}\frac{d_{\theta'}^2}{d_\theta^2} \partial_\theta \log d_{\theta'}
\end{gather*}
\begin{align*}
R^\star_{rr} =& - N\left[\partial^2_r \log f_2 - (\partial_r \log f_1) (\partial_r \log f_2) + (\partial_r\log f_2)^2\right], \qquad
R^\star_{ra} = 0\\
R^\star_{a a'} 
=&- \delta_{aa'}g^2_a \frac{f_2^2}{f_1^2}\left [N(\partial_r\log f_2)^2  + \partial_r^2 \log f_2 
-(\partial_r\log f_1)(\partial_r\log f_2)\right ] + \hat R_{aa'}
\end{align*}
($ \hat R_{a a'}$ is the Ricci tensor for the subspace spanned by variables $\theta_a$).
%& + \delta_{\theta\theta'} \sum_{\theta''} \frac{d_\theta^2}{d^2_{\theta''}} \left [(\partial_{\theta''} \log d_\theta) 
%\frac{\partial}{\partial \theta''} (2\log d_{\theta''}-\omega ) - 2\frac{\partial^2_{\theta''} d_\theta}{d_\theta} \right]\\
%& - (\partial_\theta \log d_{\theta'})(\partial_{\theta'}\log d_{\theta'}) -(\partial_\theta \log d_\theta)(\partial_{\theta'} \log %d_{\theta})
%+2(\partial_\theta \log d_{\theta'})(\partial_{\theta'} \log d_{\theta})\\
%& + \frac{1}{d_{\theta'}}\partial^2_{\theta\theta'} d_\theta + \frac{1}{d_{\theta}}\partial^2_{\theta\theta'} d_{\theta'}
%-\partial^2_{\theta\theta'}\omega +(\partial_\theta \log d_{\theta'})\partial_{\theta'}\omega
%+(\partial_{\theta'} \log d_{\theta})\partial_{\theta}\omega\\
%&+ \sum_{\theta''} (\partial_{\theta}\log d_{\theta''})(\partial_{\theta'}\log d_{\theta''})\qquad \omega=\sum_\theta \log %d_\theta\\
%
\begin{align}
R^\star 
%=& \frac{1}{f_1^2}R^\star_{rr} +\frac{1}{f_2^2} \sum_\theta \frac{1}{g^2_\theta} R^\star_{\theta\theta}\\
= \frac{\hat R}{f_2^2} - \frac{N}{f_1^2}\left[2\partial^2_r \log f_2 - 2(\partial_r \log f_1) (\partial_r \log f_2)    +(N+1) (\partial_r\log f_2)^2 \right ] \label{RstarStar}
\end{align}
If $\theta_a$ parametrize the unit sphere $S_N$, then $\hat R=N(N-1)$
 (\cite{Petersen} p.65). 
% If $N=2$:
%\begin{align*}
%R^\star = \frac{2}{f_2^2} - \frac{2}{f_1^2}\left[2\partial^2_r \log f_2 - 2(\partial_r \log f_1) (\partial_r \log f_2) 
%+3 (\partial_r\log f_2)^2 \right ]
%\end{align*}
%For $f_1=1$ and $f_2=r$: $ds^2 = dr^2 + r^2d\Omega_N$: $R^\star = \frac{1}{r^2}[N(N-1) - N(N-1)]=0 $.\\

\subsection{Space components of the electric tensor}\label{EVAL_EMUNU}
The normalization condition $u^ku_k=-1$ is $-1=-b^2(r) u^0u^0$. Then $u^0= 1/b$, $u_0=-b$. 
\begin{align*}
E_{\mu\nu} =& u^0 C_{0\mu\nu}{}^0 u_0 = - C_{0\mu\nu}{}^0 \\
=& R_{\mu 0\nu}{}^0 - \frac{R_{\mu\nu} +R_0{}^0g_{\mu\nu} }{n-2} + \frac{R g_{\mu\nu}}{(n-1)(n-2)}\\
=& -\frac{1}{b} \nabla^*_\mu b_\nu + \frac{a a_{tt}}{b^2} g^\star_{\mu\nu} \\
&- \frac{1}{n-2}\Big [ R^\star_{\mu\nu} + g^\star_{\mu\nu}  \frac{1}{b^2} [ (n-2)a_t^2  + aa_{tt} ] 
-\frac{1}{b} \nabla^\star_\mu  b_\nu 
- \frac{1}{b}\nabla^\star_\rho b^\rho g^\star_{\mu\nu} +(n-1)\frac{a a_{tt}}{b^2}g^\star_{\mu\nu} \Big]\\
& +\left[\frac{1}{a^2}R^\star -\frac{2}{a^2b}\nabla^\star_\mu b^\mu + (n-1)\frac{1}{a^2b^2}[(n-2)a_t^2 + 2a a_{tt} ]\right]
\frac{a^2 g^{\star}_{\mu\nu}}{(n-1)(n-2)}\\
=& - \frac{n-3}{n-2}\frac{1}{b} \left [ \nabla^\star_\mu b_\nu -\nabla^\star_\rho b^\rho  
\frac{g^{\star}_{\mu\nu}}{n-1}\right ]  -  \frac{1}{n-2}\left [R^\star_{\mu\nu}  - R^\star \frac{g^{\star}_{\mu\nu}}{n-1}\right].
\end{align*}

\end{document}